\documentclass[useAMS,usenatbib,usegraphicx]{mn2e}

\newcommand{\lya}{Lyman~$\alpha$}
\newcommand{\ev}{\mathrm{eV}}
\newcommand{\ang}{\mathrm{\AA}}
\newcommand{\lamal}{\lambda_\alpha}

\newcommand{\nual}{\nu_\alpha}
\newcommand{\lal}{L_\alpha}
\newcommand{\delnu}{\Delta\nu_\mathrm{D}}
\newcommand{\fesc}{f_\mathrm{esc}}
\newcommand{\vc}{v_\mathrm{c}}
\newcommand{\msun}{\mathrm{M}_\odot}
\newcommand{\sfr}{\mathrm{M_\odot\,yr^{-1}}}
\newcommand{\yr}{\mathrm{yr}}
\newcommand{\dd}{\mathrm{d}}

\newcommand{\kelvin}{\mathrm{K}}
\newcommand{\rs}{R_\mathrm{S}}
\newcommand{\rvir}{R_\mathrm{vir}}
\newcommand{\mh}{m_\mathrm{H}}
\newcommand{\tsf}{t_\mathrm{SF}}
\newcommand{\nel}{n_\mathrm{e}}
\newcommand{\nh}{n_\mathrm{H}}
\newcommand{\nhi}{n_\mathrm{HI}}
\newcommand{\ala}{\alpha_\mathrm{A}}
\newcommand{\alb}{\alpha_\mathrm{B}}

\newcommand{\xhi}{x_\mathrm{HI}}
\newcommand{\xigm}{x^\mathrm{IGM}_\mathrm{HI}}
\newcommand{\kpc}{\mathrm{kpc}}
\newcommand{\kms}{\mathrm{km\,s^{-1}}}
\newcommand{\vd}{v_\mathrm{D}}
\newcommand{\cmmt}{\mathrm{cm}^{-2}}
\newcommand{\cmt}{\mathrm{cm}^{2}}
\newcommand{\lum}{\mathrm{erg~s}^{-1}}
\newcommand{\flux}{\mathrm{erg~s^{-1}~cm^{-2}}}
\newcommand{\fv}{f_v}
\newcommand{\jto}{J_{21}}
\newcommand{\hi}{{\mbox{H\,{\sc i}}}}
\newcommand{\hii}{{\mbox{H\,{\sc ii}}}}

\newcommand{\heii}{{\mbox{He\,{\sc ii}}}}
\newcommand{\pmpc}{\mathrm{pMpc}}
\newcommand{\vs}{v(\rs)}
\newcommand{\phil}{\phi_\lambda}
\newcommand{\val}{v_{\alpha}}
\newcommand{\vvir}{v(\rvir)}

\title[Probing reionization with Lyman~$\alpha$ emission
lines]{Probing reionization with Lyman~$\alpha$ emission lines}

\author[M. R. Santos]{Michael R. Santos\thanks{E-mail:
mrs@tapir.caltech.edu}\\ Theoretical Astrophysics, California
Institute of Technology, Mailcode 130-33, Pasadena, CA 91125, USA}

\begin{document}

\date{\today}

\pagerange{\pageref{firstpage}--\pageref{lastpage}} \pubyear{2003}

\maketitle

\label{firstpage}

\begin{abstract}
\lya\ emission from high-redshift galaxies may be a powerful probe of
the ionization history of the IGM at $z>6$: the observed \lya\
emission line is sensitive to the neutral fraction of IGM hydrogen in
the range 0.1--1.  We present calculations of observed \lya\ emission
lines from $z>6$ galaxies, illustrating the effect of varying the many
free parameters associated with the emitting galaxy, its halo, and the
IGM around the galaxy.  In particular, we use a dynamic model of the
IGM that includes the effect of IGM infall toward the emitting galaxy.
Galactic winds may play a crucial role in determining observed \lya\
line fluxes.  We compare our model predictions with observations of
two $z=6.5$ galaxies and conclude that, if galactic winds are allowed
for, existing observations place no constraint on the neutral fraction
of the IGM at $z=6.5$.  Future space-based observations will constrain
the importance of galactic winds; if winds are unimportant for the
observed $z=6.5$ galaxies, our models suggest that the IGM neutral
fraction at $z=6.5$ is $\la0.1$.
\end{abstract}


\section{Introduction}

Strong \lya\ emission is present in many distant galaxies, and is an
important signpost for discovering high-redshift galaxies and,
especially, measuring redshifts for them.  The presence of strong
\lya\ emission has been crucial in confirming the highest redshift
galaxies known at $z>5$
\citep[e.g.,][]{hu99,hu02a,hu02b,aji03,kod03,rho03}.  See
\citet{tan03} for a review.  It also plays an important role in
identifying the redshift of some galaxies at $z\ga2$ because its
strength, even in some apparently extremely dusty galaxies, allows for
redshift determinations when little continuum light is visible
\citep[e.g.,][]{ste03,cha03}.

In this paper we concentrate on \lya\ emission powered by star
formation.  Hot stars emit photons capable of ionizing hydrogen.  If
the ionized gas density is sufficient, hydrogen will recombine on
relevant timescales.  The recombination cascade produces a rich
spectrum of lines, but usually includes a \lya\ photon \citep[see][for
a review]{ost89}.  Because most recombinations produce \lya\ photons,
it is predicted to be the strongest line emitted from recombination
emission. AGN activity can also provide ionizing photons that
ultimately generate a strong \lya\ line.  Additional sources of \lya\
photons are atomic-hydrogen cooling of gas
\citep[e.g.,][]{hai00,far01} and fluorescence of gas clouds
illuminated by a strong, non-local source of ionization
\citep[e.g.,][]{reu03}.

The predicted strength of the \lya\ line, and the assumed prevalence
of star formation at early times (inferred from the abundance of old
stars in the local universe) led to an early prediction of an abundant
population of high-redshift galaxies with very strong \lya\ emission
lines \citep{par67}.  Though subsequent revisions to the theory of
galaxy formation, primarily that it was a hierarchical process,
explained the relative lack of \lya\ detections compared to those
early predictions \citep{hai99}, searching for galaxies based solely
on strong \lya\ emission has been validated recently as an effective
technique for identifying galaxies up to $z=6.5$.

\lya\ is a resonant transition with a large cross-section.  As a
consequence, small quantities of neutral hydrogen scatter away \lya\
photons in direction and frequency.  In particular, if the
intergalactic medium (IGM) is not highly ionized, as is expected for
at least some of the time between recombination ($z\sim1100$) and
$z\simeq6$ \citep{fan02,kog03}, then the neutral hydrogen in the IGM
can easily scatter the \lya\ line from galaxies.  The \lya\ photons
are not destroyed, but the scattering process diffuses the line in
areal coverage and frequency \citep{loe99}, and the resulting low
surface brightness emission is currently unobservable.

A simple consideration of the IGM shows that the blue side of the
emergent \lya\ line would be scattered away by a neutral IGM, and only
the red side of the line would be observed.  This follows from
considering that most of the scattering by neutral hydrogen comes at
wavelengths near the \lya\ transition in the rest-frame of the gas.
However, for a completely neutral IGM at high redshifts, the \lya\
cross-section far from resonance becomes important, since the natural
line profile has `damping wings' that fall off only like $\nu^{\pm2}$.
Thus even if the IGM has no component at the velocity of corresponding
to resonant scattering of a particular \lya\ photon, that photon may
still be scattered.  This observation led to the conclusion that
observations of \lya\ emission lines (or their absence) at high
redshifts may probe the ionization state of the IGM in a regime poorly
tested by the Gunn-Peterson trough measurements, $\xhi\sim1$
\citep{mir98a,mir98b,hai02,bar03b,cen03b}.

In particular, since the \textit{WMAP} satellite discovered a high
optical depth to the last scattering surface \citep{kog03}, the
ionization history of the universe at $z>6$ is of great interest.
Though the \textit{WMAP} results indicate that the universe was mostly
ionized for a substantial history of the universe at $z>6$, it does
not constrain the exact ionization level or the history in detail.
There are now many predictions for the ionization history at $z>6$
based on semi-analytic and numerical modelling of early star formation
\citep[e.g.,][]{cen03a,wyi03,cia03,hai03}.  If there is a population
of strong \lya\ emitters at the redshifts in question, $6\la z\la20$,
and if the effect of IGM scattering on their observed properties can
be calculated, then future surveys for \lya\ emission may provide
valuable information on the reionization history at $z>6$.

Here we use a more sophisticated treatment of the dynamics of the IGM
around galaxies with strong intrinsic \lya\ emission to investigate
the effect of the IGM on the observed line.  The IGM model we use will
be presented in more detail separately, but is summarized here and
nearly identical to that described in \citet{bar02}.  The key feature
is that infall of the IGM toward the galaxies is included.
\citet{bar03a} have showed that such an IGM model makes a prediction
for an absorption feature in the spectra of high-redshift QSO \lya\
lines.  Here we consider normal galaxies, where the effect of an
infalling IGM is even stronger.

Throughout this paper we assume a cosmology based on the recent
results from \textit{WMAP} and other work \citep[and references
therein]{spe03}: $\Omega_\mathrm{M}=0.3$, $\Omega_\Lambda=0.7$,
$H_0=70~\kms$, and $\Omega_\mathrm{b}=0.0469$.  In general we quote
distances in physical Mpc, denoted by `pMpc.'

This paper is organized as follows.  In Section~\ref{sec:lyascat} we
present the \lya\ scattering cross-section.  Section~\ref{sec:lyaprof}
describes the model for the intrinsic \lya\ line profile and the
observed profile modified by IGM scattering.  Section~\ref{sec:lyares}
presents \lya\ line profiles as a function of the (many) input
parameters that influence the observed line profile.  We introduce the
possible role of galactic winds in Section~\ref{sec:galwind}.
Section~\ref{sec:disc} discusses our \lya\ flux predictions with
comparison to current observations, and also reviews model
assumptions.  Section~\ref{sec:summ} summarizes.

\section{Lyman~$\alpha$ scattering}
\label{sec:lyascat}

A \lya\ photon is emitted when a hydrogen atom makes a transition from
the $n=2$ level to the $n=1$ ground-state.  The emitted photon has an
energy of $10.199~\ev$ and a wavelength of $\lamal=1215.67~\ang$.  The
transition is the strongest spontaneous transition of a hydrogen atom,
with an Einstein $A$ value of
$A_{21}=6.265\times10^8~\mathrm{s}^{-1}$.  Correspondingly, the
absorption cross-section of the transition is relatively large as
well.

A \lya\ photon travelling through a region of neutral hydrogen will be
repeatedly absorbed and re-emitted.  This process by itself does not
destroy \lya\ photons, but if photons are scattered over a
sufficiently large region such that an observation spatially resolves
the emission, then the \lya\ line will be difficult to observe in
practice \citep{loe99}.  In this work we assume that \lya\ photons
scattered in the IGM, that is, outside of the virial radius of the
emitting galaxy, are missed by observation.  For a $z\ga6$ galaxy, the
virial radius is larger than 1\arcsec\ (the size of a typical
spectroscopic slit) for any galaxy more massive than
$\sim3\times10^9~\msun$.

So under the assumption that photons emitted in the \lya\ emission
line of a galaxy are lost from subsequent observation if they are
scattered by the IGM, we constructed a model of the IGM (see
Section~\ref{sec:igmmodel}) and computed the absorption due to neutral
hydrogen.  The natural absorption cross-section $\sigma_\mathrm{N}$ is
\citep{pee93}
\begin{equation}
\sigma_\mathrm{N}(\nu) = \frac{3 \lamal^2 A_{21}^2}{8\pi} 
\frac{(\nu/\nual)^4}{4\pi^2(\nu-\nual)^2 + (A_{21}^2/4)(\nu/\nual)^6},
\end{equation}
where $\nu=c/\lambda$ and $\nual=c/\lamal$.

\begin{figure}
  \includegraphics[width=8.4cm]{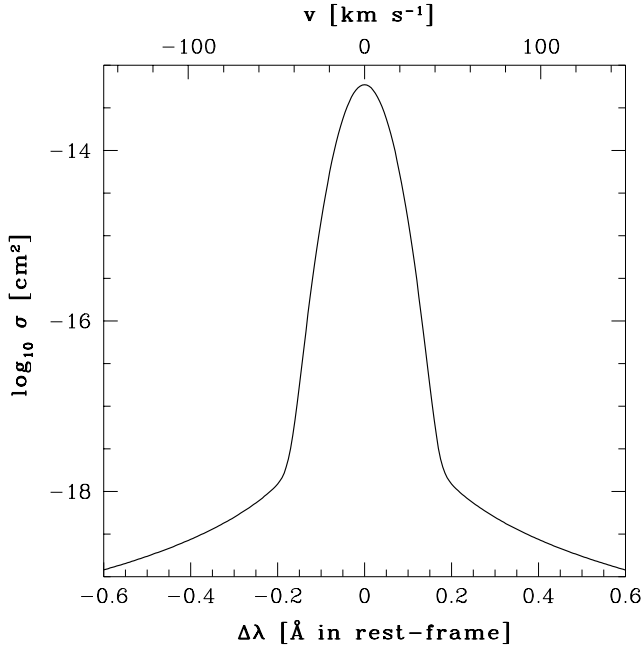}
  \caption{Cross-section for \lya\ absorption, as a function of
  wavelength difference from resonance in the frame of the absorbing
  atom.  The top axis shows the velocity (in the frame of the
  absorbing atom) where the \lya\ resonance corresponds to the
  wavelength on the bottom axis.}
  \label{fig:lgdpxv}
\end{figure}

The total absorption cross-section due to a parcel of IGM gas is
determined by the kinetic properties of the gas, both bulk and
thermal, as well as the natural cross-section.  Bulk motions just
introduce a frequency shift, but thermal motions broaden the profile,
modelled as the convolution of the natural cross-section with a
Maxwellian velocity distribution,
\begin{equation}
\sigma_\mathrm{V}(\nu) = \int_{-\infty}^\infty 
M(v) \sigma_\mathrm{N}(\nu-\nual v/c) \dd v,
\end{equation}
\begin{equation}
M(v) = \left(\frac{\mh}{2 \pi k T}\right)^{1/2} 
\exp \left(-\frac{\mh v^2}{2 k T}\right),
\end{equation}
where $\mh$ is the mass of the hydrogen atom and $T$ is the
temperature of the IGM gas (modelled in the next section).  For the
usual Lorentzian approximation of $\sigma_\mathrm{N}$, the convolved
cross-section would be a Voigt profile, hence we label our convolved
cross-section $\sigma_\mathrm{V}$.  Figure \ref{fig:lgdpxv} shows
$\sigma_\mathrm{V}(\lambda)$, assuming a temperature of
$T=10^4~\kelvin$ for the scattering gas.  The profile has a Doppler
core with a half-width of $\sim10~\kms$ and power-law tails called
`damping wings.'  If the IGM at the redshift of interest is or has
previously been reionized, then $T=10^4~\kelvin$ is probably a good
approximation.  If the IGM is still thermally pristine, then the
temperature may be as low as $\sim20~\kelvin$, resulting in a narrower
Doppler core of the scattering profile.  In practice this would have a
small effect on any near-future observation.

\section{Lyman~$\alpha$ line profiles}
\label{sec:lyaprof}

We will refer to two types of \lya\ lines from galaxies.  The
\textbf{intrinsic line} is the line produced in the galaxy,
specifically, the line that reaches the virial radius of the halo
containing the emitting galaxy.  The \textbf{observed line} is the
line after propagation though the IGM to $z=0$.

\subsection{Intrinsic \lya\ emission line}
\label{sec:inline}

A \lya\ emission line from a high-redshift galaxy is typically a
consequence of hydrogen recombinations in nebular gas.  The ionization
of the \hii\ region may be maintained either by hot stars or by gas
accretion onto a black hole.  Though luminous QSOs have been
discovered at $z>6$ \citep[and references therein]{fan03}, deep x-ray
data indicates that most galaxies at those redshifts do not contain
powerful AGNs \citep{barg03}.  Thus in this work we will assume recent
($\la10$~Myr ago) star formation powers \lya\ emission lines.

Type O stars emit hydrogen-ionizing photons.\footnote{Approximately
1~Gyr after the onset of star formation hot white dwarfs form, and
these may emit substantial ionizing radiation \citep{cha00}; however,
for $z\ga5.5$ the universe is not old enough to have formed these
stars.}  Some fraction, $\fesc$, of these photons may escape the
galaxy without absorption.  The rest of the photons are absorbed by
either neutral hydrogen or dust within the galaxy.  An ionizing photon
absorbed by a hydrogen atom initiates a cascade of recombination
emission, ultimately resulting in a \lya\ photon about 2/3 of the
time, and two-photon emission the other 1/3 of the time \citep{ost89}.
A \lya\ photon subsequently scatters through absorption and reemission
until the optical depth of the galaxy at its wavelength is $\la1$; the
optical depth is determined both by the position and direction of the
photon, and also its wavelength.  If dust is present, it may absorb
photons at any time in this process.

The radiative transfer problem for \lya\ depends sensitively on the
geometry of neutral hydrogen and dust within the galaxy \citep{neu91},
and also the dynamics of the hydrogen \citep{kun98}.  For simplicity
we will assume that the intrinsic \lya\ emission line is centered at
the systemic redshift of the emitting galaxy, with a Doppler profile
shape described by a characteristic velocity, $v_\mathrm{D}$,
\begin{equation}
\phi_\nu^\mathrm{in}(\nu) = \frac{1}{\sqrt{\pi}\delnu} 
\exp\left[\frac{-(\nu-\nual)^2}{(\delnu)^2}\right],
\end{equation}
where $\phi_\nu^\mathrm{in}$ is the line profile per unit frequency
and
\begin{equation}
\delnu = \frac{v_\mathrm{D}}{c}\nual.
\end{equation}
In Section~\ref{sec:galwind} we address other possible intrinsic line
profiles.

In this simplified model, two parameters characterize an intrinsic
\lya\ emission line: its luminosity, $\lal$, and $v_\mathrm{D}$.
$\lal$ depends on the rate of production of ionizing photons,
$\dot{Q}$, the ionizing photon escape fraction, $\fesc$, and dust
absorption.  The production of ionizing photons in turn depends on the
star-formation rate and the metallicity and initial mass function
(IMF) of the stars formed.  None of these quantities may be reliably
predicted from first principles for a galaxy at high redshift.
Therefore, any use of the \textit{observed} \lya\ line of a
high-redshift galaxy to deduce the radiative transfer properties of
the IGM will be complicated, unless these quantities can be measured
independently of the \lya\ line.

The estimation of $v_\mathrm{D}$ is simpler.  The likely minimum
velocity scale for motion in a galaxy is the circular velocity of the
parent halo, $\vc$; this velocity may describe the gas motion is the
star-formation is mediated by a violent process, such as the merging
of two equal-mass galaxies.  If star-formation is a more quiescent
process in the galaxy, it may occur in a disk whose peak velocity is,
roughly, between $\vc$ and $2\vc$, for realistic halo and disk
properties \citep{mo98,col00}.  Thus we assume $\vc\la v_\mathrm{D}\la
2\vc$.

\subsection{Observed \lya\ emission line}

The optical depth for photons to be absorbed in the IGM is
\begin{equation}
\tau(\nu) = \int_{\rvir}^{R_\mathrm{obs}} \xhi(r)\nh(r)
\sigma_\mathrm{V}(\nu')\dd r,
\end{equation}
\begin{equation}
\nu' = \nu\left[1+\frac{v(r)}{c}\right],
\label{eq:tau}
\end{equation}
where $R_\mathrm{obs}$ is the physical distance to the observer
(though there is almost no contribution from $r\gg\rvir$ because the
Hubble flow eventually redshifts the IGM \lya\ resonance far from the
emitted \lya\ line [in the frame of the galaxy]).  We describe our
model for $\nh(r)$, $v(r)$ and $\xhi(r)$ in
Section~\ref{sec:igmmodel}.

The observed \lya\ emission line, in the rest-frame of the emitting
galaxy, is then
\begin{equation}
\phi_\nu^\mathrm{obs}(\nu) = 
\phi_\nu^\mathrm{in}(\nu) \exp\left[-\tau(\nu)\right].
\end{equation}

\subsubsection{IGM model}
\label{sec:igmmodel}

\paragraph{Density and velocity}

We will present a full description of our model for the density and
velocity of IGM material in another paper (Santos \& Adelberger, in
preparation).  The model is very similar to one described in
\citet{bar02}.  Here we summarize our model as it applies to this
paper.

The evolution of the density and velocity of the IGM follow from the
small, linear, initial perturbations imprinted on the universe at very
high redshift.  We model the IGM starting from the statistical
description of linear perturbations by applying a simple description
of the non-linear evolution of the perturbations.

Specifically, we start by considering a \lya\ emitting galaxy in a
halo of total mass $M$ at redshift $z$.  We then constrain the initial
linear overdensity on the scale of $M$ to produce the halo at redshift
$z$.  Starting from that initial linear overdensity constraint, we use
the excursion set formalism \citep{bon91,lac93} to compute the
mean\footnote{i.e., averaged over a large ensemble of identical halos}
initial linear overdensities, averaged in spheres centered on the
location of the halo, on scales larger than the scale of $M$.  The
mean initial linear overdensity decreases monotonically with
increasing scale; this is not externally imposed.  Our approach also
implies that the halo is not substructure of a larger collapsed halo
(see also \citealt{bar02}).

Once we have the initial linear overdensities of spheres enclosing the
halo, we evolve those overdensities using the spherical top-hat model
\citep{par67}.  Because the mean initial linear overdensity decreases
monotonically with increasing scale, there is no shell crossing,
and the evolution of each IGM matter shell is independent of the
evolution of the other shells.

Our dynamical evolution treats all matter equally, and we assume the
baryons trace the dark matter in these shells.  This description would
be inappropriate if, e.g., the \lya\ emitting galaxy blows a strong
wind into its surrounding IGM \citep[see][for evidence of this at
$z=3$]{ade03}, and we will return to this in
Section~\ref{sec:galwind}.

A realistic distribution of initial linear overdensities around a
halo, in contrast to the mean initial linear overdensities averaged
over many halos, would not in general be spherically symmetric or even
monotonic with radius within spherically-averaged shells.  In addition
to collapsing toward the galaxy, there would be structures collapsed
and collapsing within the `IGM.'  The most straightforward way to
assess the effect of these complicated dynamics would be with numerical
simulation.  Here we appeal to analytic results to demonstrate the
effect of structure with the collapsing IGM will not severely affect
our conclusions.  

\citet{sca02} computed the bivariate probability distributions of two
halos forming at different positions.  Their results suggest that
outside the virial radius of a halo, the typical enhancement in the
number density of halos is, at maximum, only a factor of a few, for
any mass.  Thus we conclude that the fraction of matter collapsed into
halos (above a given mass) in the IGM surrounding our galaxy of
interest is at most a few times the universal collapse fraction.  At
$z=6$ up to half the mass of the IGM may be in the form of collapsed
halos with virial temperatures above $10^4~\kelvin$, that is, halos
massive enough to collisionally ionize their hydrogen.  That fraction
decreases with increasing radius.  The enhancement is also a
decreasing function of decreasing mass.  Thus we conservatively expect
that our simple picture gives at most a factor of two overestimate of
the optical depth, averaged over a reasonable sample of halos.  At
large optical depths, this error doesn't affect the observed \lya\
line properties, and at low optical depths we may underestimate the
observed line by up to that same factor of two.

The practical output of our model is, given a halo mass $M$ at
redshift $z$, the radius, velocity, and mass of shells of IGM matter
surrounding the galaxy from its virial radius to a large radius.

\paragraph{Ionization state}

To calculate the scattering of a \lya\ emission line by the IGM, we
need to know the ionization state of the hydrogen, in addition to its
density and velocity profile.  We solve the ionization balance of the
IGM gas assuming two possible contributions, direct ionization by the
galaxy emitting the \lya\ line and, if the universe is ionized, a mean
ionizing background.

After calculating the density distribution around a halo using the
prescription described in the previous section, we solve the
ionization balance as a function of distance from the galaxy.  In
general, the recombination rate of ionized hydrogen in the IGM is
relatively small.  As a consequence, in the pre-reionized universe
\hii\ regions around galaxies are not in equilibrium, but began
growing when star-formation turned on and expand at the rate that
ionizing photons can ionize more neutral IGM (rather than balancing
recombinations within the ionized region).  The \hii\ region is still
expected to have a relatively sharp boundary, as the mean free path of
ionizing photons in the neutral IGM is small.  After reionization, all
of the IGM is highly ionized by the mean ionizing background, but the
direct ionizing flux of the galaxy provides an additional proximity
effect on the ionization balance immediately around it.

We divide the IGM around into one or two regions: either the universe
is reionized, in which case all of the IGM is ionized; or the universe
is not reionized, in which case there is an \hii\ region in the
IGM immediately surrounding the galaxy, and the IGM is fully neutral
outside of that region.  In that case the radius $\rs$ of the
\hii\ region is calculated by solving
\begin{equation}
\int_{\rvir}^{\rs} \frac{4 \pi r^2 \nh(r)}{\dot{Q}-\Lambda(r)}\dd r =
\tsf,
\end{equation}
\begin{equation}
\nh(r) = \frac{\left[ \delta(r)+1 \right] \rho_{\mathrm{H},0}(1+z)^3}{\mh},
\end{equation}
\begin{equation}
\dot{Q} = \int_{\nu_0}^\infty \frac{L_\nu}{h \nu} \dd \nu,
\end{equation}
\begin{equation}
\Lambda(r) = \int_{\rvir}^{r} 4 \pi \tilde{r}^2 \nh^2(\tilde{r}) \alb(T) 
\dd \tilde{r},
\end{equation}
where $\rvir$ is the virial radius of the galaxy's halo, $\tsf$ is the
age of the star-formation activity\footnote{We model star formation
with a constant rate for an age $\tsf$.}, $\delta(r)$ is the
overdensity with respect to the mean IGM density (calculated with the
model in the previous section), $\rho_{\mathrm{H},0}$ is the comoving
(i.e., $z=0$) density of hydrogen, $z$ is the redshift of the galaxy,
$h\nu_0$ is the ionization energy of hydrogen, $L_\nu$ is the specific
luminosity of the galaxy, and $\alb(T)$ is the hydrogen Case B
recombination coefficient.  Since the \hii\ region is almost
completely ionized, here we assumed the electron physical density,
$\nel$, is $\nel=\nh$, and ignored the contribution to $\nel$ from
helium.  We have not included a clumping factor modification to
the recombination rate; however, we do account for the mean
overdensity of the IGM matter.  Additional clumping of this material
would decrease the size of the true \hii\ region compared to our
calculation.

The ionization rate for a hydrogen atom within an ionized region a
distance $r$ from the galaxy is
\begin{equation}
\Gamma(r) = \int_{\nu_0}^\infty 
\left(\frac{L_\nu}{4 \pi r^2} + 4 \pi J_\nu \right)
\frac{\sigma_\mathrm{V}(\nu)}{h \nu} \dd \nu,
\label{eq:ionrate}
\end{equation}
where $J_\nu$ is the mean specific intensity of the intergalactic
background radiation.  If the universe is not reionized, then we set
$J_\nu=0$.  The hydrogen neutral fraction at any radius, $\xhi(r)$, is
solved for by setting $\Gamma(r)$ equal to the recombination rate at
that radius,
\begin{equation}
\xhi(r) = 1 + \frac{\chi}{2} - \frac{1}{2} \left(\chi^2+4\chi\right)^{1/2}
\end{equation}
\begin{equation}
\chi = \frac{\Gamma(r)}{\nh(r) \ala(T)},
\end{equation}
where $\ala(T)$ is the hydrogen Case A (total) recombination
coefficient.  Additional clumping of the IGM would decrease $\chi$
linearly, increasing $\xhi$.

We assume that when IGM gas falls across the virial radius it shocks
and ionizes completely, and has no impact on the \lya\ line profile.

\section{Lyman~$\alpha$ line results}
\label{sec:lyares}

As described in Section~\ref{sec:inline}, the strength and shape of
the intrinsic \lya\ emission line depends on many properties of the
source galaxy.  Additionally, the properties of the observed line are
strongly modified by scattering in the IGM, which depends on the
properties of the galaxy, its halo, and the IGM model.  To illustrate
these dependencies, we will adopt a fiducial model for a \lya\
emitting galaxy, then show the dependence of the intrinsic and
observed \lya\ line on the properties of that galaxy and the IGM.

\subsection{Fiducial galaxy/halo model}

Our fiducial \lya\ emitter model is a galaxy at rest in the center of
a $10^{11}~\msun$ halo at $z=6.5$.  `Fiducial galaxy' will refer to
this total system, galaxy plus halo.  We assign the fiducial galaxy a
constant star-formation rate (SFR) of $10~\sfr$ for an age of
$\tsf=10^8~\yr$.  The virial radius of the halo is $\rvir=19.8~\kpc$,
and the circular velocity at the virial radius is $\vc=148~\kms$.  The
halo dynamical time is then $1.3\times10^8~\yr$.  Given our cosmology,
and assuming the universal baryon to dark matter ratio holds in the
halo, the halo contains $1.5\times10^{10}~\msun$ of baryons.  Thus at
$10~\sfr$ the gas supply could last up to $1.5\times10^{9}~\yr$, about
twice the age of the universe at that redshift.  The chosen SFR also
implies that 9 per cent of the baryons are converted into stars per
halo dynamical time.

We use the following prescription to convert the fiducial galaxy
properties into the emitted \lya\ line.  We convert the SFR to a
hydrogen-ionizing photon luminosity using
\begin{equation}
\dot{Q} = 3.5\times10^{53}~\mathrm{s^{-1}} 
\left(\frac{\mathrm{SFR}}{\sfr}\right),
\end{equation}
appropriate for a Salpeter IMF from 1 to $100~\msun$ with $1/20$th
solar metallicity \citep{lei99,sch03}.  We assume that $2/3$ of the
ionizing photons that do not escape are converted into \lya\ photons
\citep{ost89}, so that
\begin{equation}
\lal = \frac{2}{3} (1-\fesc) \dot{Q} h \nual.
\end{equation}
Note that this assumes there is no absorption by dust; we return to
the issue of dust in Section~\ref{sec:modass}.  The other \lya\ line
parameter that needs to be specified is the line width, characterized
by $\vd\equiv\fv\vc$.  For our fiducial model, we set $\fv=1.5$.

Additionally we need to model the effect of photoionization of the
galaxy on the surrounding IGM.  We assume the escaping photon
luminosity of ionizing photons is simply
\begin{equation}
\dot{Q}_\mathrm{esc} = \fesc \dot{Q}.
\end{equation}
Our baseline choice for the escape fraction is $\fesc=0.1$, giving
\begin{equation}
\lal = 3.4\times10^{42}~\mathrm{erg~s^{-1}} 
\left(\frac{\mathrm{SFR}}{\sfr}\right).
\end{equation}

\subsection{IGM model dependence}
\label{sec:igmmoddep}

Now we illustrate the IGM scattering of the \lya\ line emitted by our
fiducial galaxy, highlighting the importance of using our dynamic IGM
model.  At $z=6.5$, the ionization state of the IGM is unknown.
Limits from QSO spectra at redshifts approaching 6.5 indicate that
$\xhi>10^{-3}$ averaged over the volume of the IGM, and $\xhi>10^{-2}$
averaged over the mass of the IGM \citep{fan02}.  Simulations of
reionization suggest that the $z\sim6$ transition from $\xhi\simeq1$
to $\xhi\simeq0$ was very sudden \citep{gne00}.  Combined with the QSO
observations, these suggest that $\xhi$ could have almost any value at
$z=6.5$.  The \textit{WMAP} results constrain the IGM ionization
fraction to be $\sim1$ over a substantial fraction of the evolution of
the universe at $z>6$, but do not provide detailed constraints on
models of $\xhi(z)$.

We start by assuming our fiducial galaxy ($z=6.5$) is embedded in a
fully neutral IGM.  \citet{hai02} argued that the detection of
reasonably strong \lya\ emission from $z=6.5$ galaxies did not rule
out a fully neutral IGM, as had been previously suggested
\citep{hu02a,rho03}.  However, the calculations of \citet{hai02} did
not account for the infall profile of the IGM around a $z=6.5$ galaxy.

\begin{figure}
  \includegraphics[width=8.4cm]{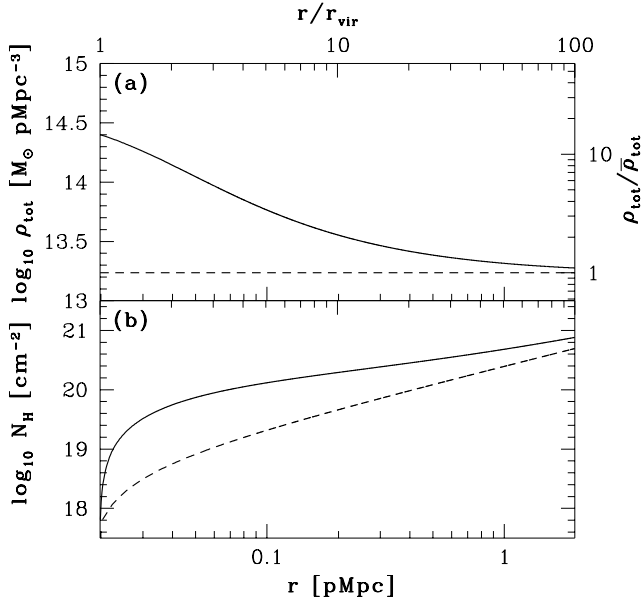}
  \caption{IGM density as a function of radius from the fiducial
  galaxy.  The left edges of the plots are at the virial radius. (a)
  The total IGM mass density and density contrast as a function of
  radius in physical Mpc.  (b) The cumulative IGM hydrogen column
  density to radius $r$.  In both plots the solid curve is for the
  dynamic IGM model and the dashed curve is for the simple model.}
  \label{fig:lgrlgrholgnh_fiduc_fidi1}
\end{figure}

First we illustrate the difference between our model of the IGM, and a
smooth, Hubble flow model.  We call our model the \textbf{dynamic IGM
model} because we allow the IGM to evolve away from the comoving
solution.  We call the mean-density, comoving model the \textbf{simple
IGM model}.  The simplest difference between the models is the density
of the IGM as a function of distance from the center of the halo.
Figure \ref{fig:lgrlgrholgnh_fiduc_fidi1}a shows the density and
density contrast of the IGM outside the halo virial radius for our
dynamic IGM model, and a simple IGM.  Within about 10 virial radii, or
0.2 physical Mpc (pMpc), the density enhancement is between 2 and 10
times the mean IGM density.  Integrating the density out from the
virial radius gives the curves shown in
Fig. \ref{fig:lgrlgrholgnh_fiduc_fidi1}b, after converting total
density to hydrogen density.  Though
Fig. \ref{fig:lgrlgrholgnh_fiduc_fidi1} shows that there is a
substantial difference in the density profile at small radius, the
increased density as a function of radius is not the most important
effect of the dynamic IGM model.

\begin{figure}
  \includegraphics[width=8.4cm]{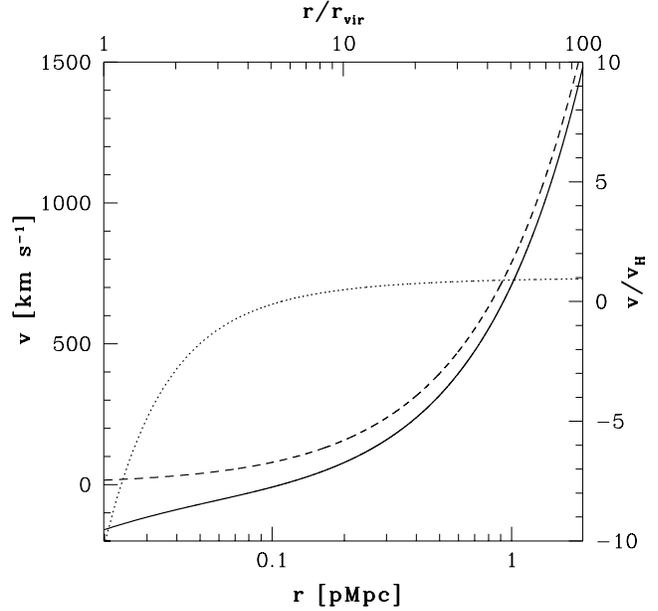}
  \caption{Velocity of IGM gas as a function of radius from the
  fiducial galaxy.  The solid curve is the dynamic IGM model; the
  dashed curve is the Hubble flow relation of the simple model.  The
  dotted curve is the ratio of the dynamic model to the simple model,
  in units marked on the right axis.}
  \label{fig:velrad_fiduc_fidi1}
\end{figure}

\begin{figure}
  \includegraphics[width=8.4cm]{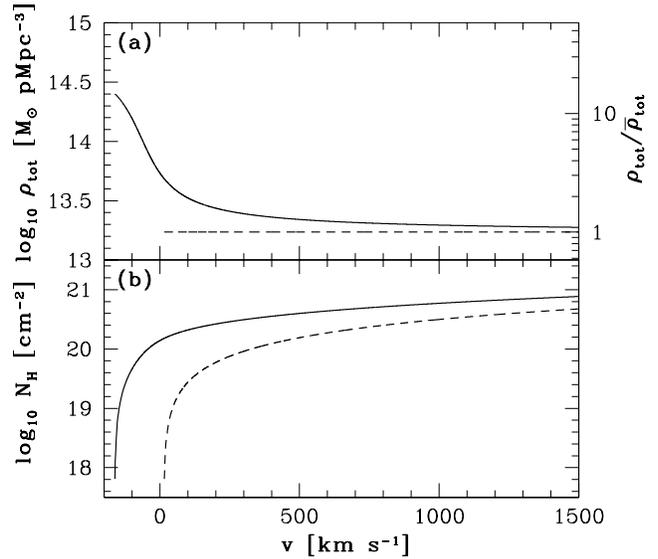}
  \caption{IGM density as a function of velocity in the frame of the
  fiducial galaxy.  (a) The total IGM mass density and density
  contrast as a function of velocity.  (b) The cumulative IGM hydrogen
  column density to velocity $v$.  In both plots the solid curve is for
  the dynamic IGM model and the dashed curve is for the simple model.}
  \label{fig:vlgrholgnh_fiduc_fidi1}
\end{figure}

The most important difference between the IGM models, with respect to
scattering of \lya\ line photons, is the distribution of density as a
function of velocity.  Figure \ref{fig:velrad_fiduc_fidi1} shows the
relationship between radius and velocity, for the dynamic and simple
IGM models.  With this relation we replot the densities of
Fig. \ref{fig:lgrlgrholgnh_fiduc_fidi1} against velocity in
Fig. \ref{fig:vlgrholgnh_fiduc_fidi1}.  The local Hubble parameter is
$H(z=6.5)=790~\kms~\pmpc^{-1}$.  For the IGM at positive velocities
with respect to the galaxy, that is, the blue side of the line, there
is little difference in density between the dynamic and simple models.
However, while the simple model has no gas at negative velocities,
corresponding to the red side of the line, the dynamic model has
reached a column density of hydrogen of $10^{20}~\cmmt$ by the time
the velocity has reached 0.  The absorption profile
(Fig.~\ref{fig:lgdpxv}) has a cross-section of greater than
$5\times10^{-14}~\cmt$ over the central $10~\kms$.  Thus the average
optical depth within about $100~\kms$ of the red side of the line is
greater than $5\times10^4 \xhi$.  Immediately we can see that for
$\xhi\sim10^{-5}$, a typically expected value for the IGM, optical
depths of order unity and greater may be expected.  This optical depth
on the red side of the observed line due to scattering by highly (but
not completely) ionized gas gives rise to a line profile distinctly
different from a simple IGM model.

\begin{figure}
  \includegraphics[width=8.4cm]{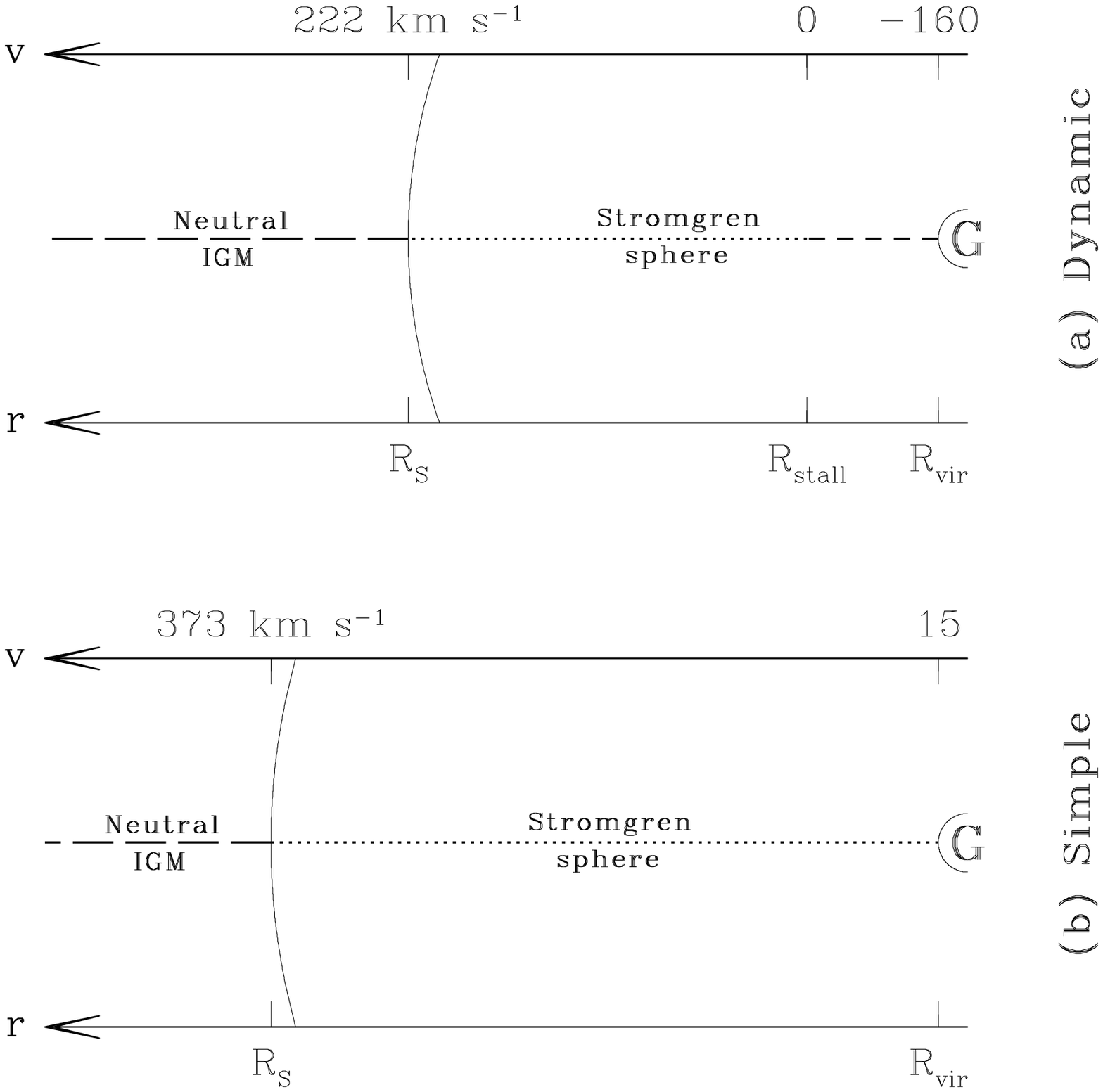}
  \caption{Schematic diagram of location and velocity of the IGM
  around the fiducial galaxy.  (a) Dynamic IGM model.  (b) Simple
  model.  The galaxy is marked by a `G' at right.  The curve
  immediately surrounding it shows the virial radius of the halo.  The
  next curve, at $\rs$, shows the Stromgren sphere.  The velocity axis
  marks the velocities (in $\kms$) corresponding to these radii.  In
  (a) $R_\mathrm{stall}$, the radius where IGM gas has zero velocity
  with respect to the galaxy, is also marked.  The dashed and dotted
  lines key the physical location of the IGM material to the
  contribution it makes to the scattering optical depth, shown in
  Fig.~\ref{fig:lamtau2_fiduc_fidi1}.}
  \label{fig:scheme}
\end{figure}

\begin{figure}
  \includegraphics[width=8.4cm]{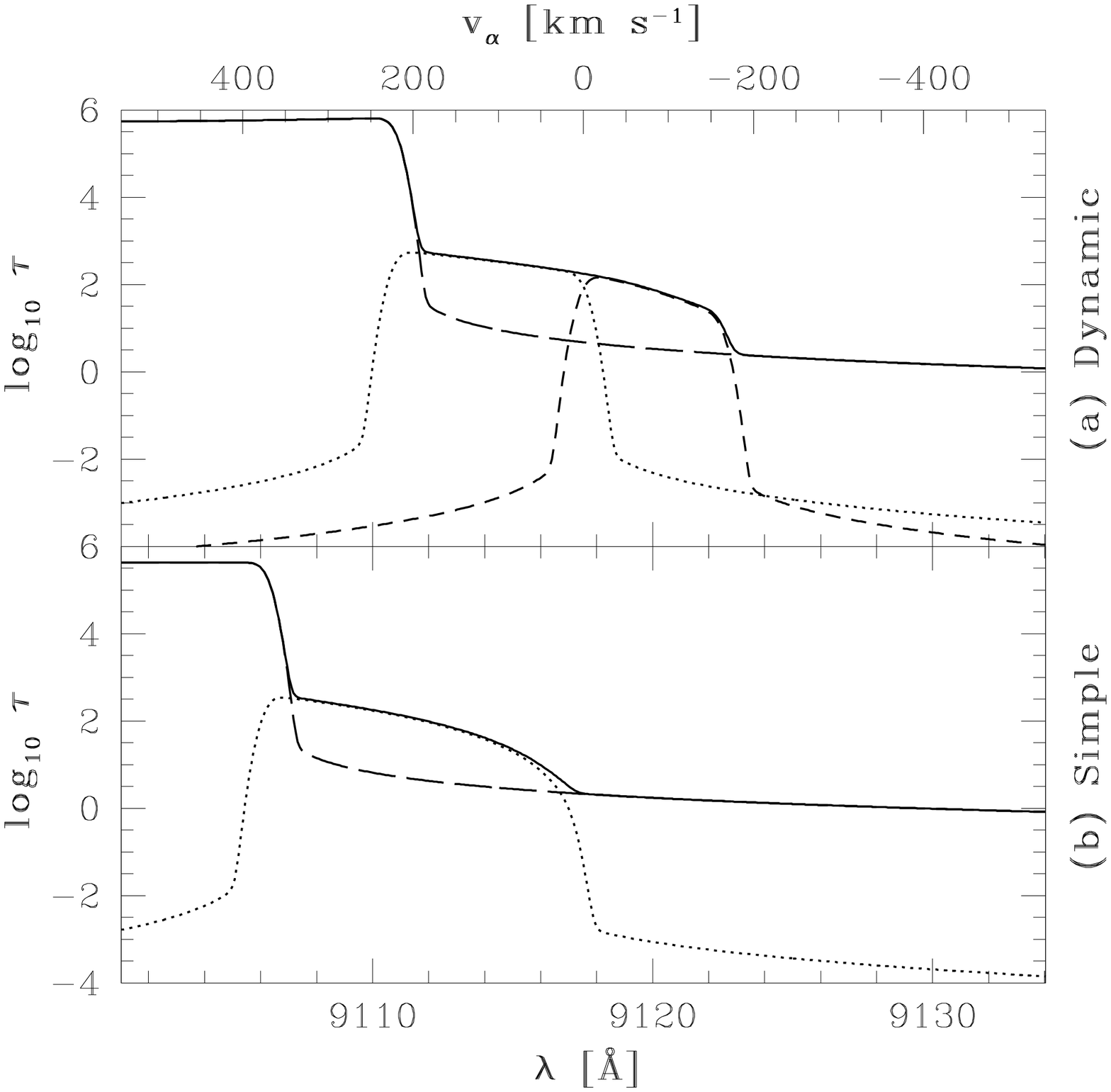}
  \caption{Optical depth due to IGM scattering as a function of
  observed wavelength.  (a) Dynamic IGM model.  (b) Simple model.  The
  solid curves are the total optical depth due to all IGM \hi.  The
  short-dashed curves are the optical depth due to gas with negative
  velocity (in the frame of the galaxy).  The dotted curves are the
  optical depth due to gas with positive velocity inside the Stromgren
  sphere.  The long-dashed curve is the optical depth due to gas
  outside the Stromgren sphere.}
  \label{fig:lamtau2_fiduc_fidi1}
\end{figure}

Figure~\ref{fig:scheme} presents a schematic illustration of the
difference between the dynamic and simple IGM models.  As described
above, we see that the dynamic model has gas with negative velocity,
in contrast to the strictly positive velocity IGM in the simple model.
We also see that the Stromgren radius is both larger and extends to
larger IGM velocities.  In Fig.~\ref{fig:scheme} we have keyed
different physical regions with different line types;
Fig.~\ref{fig:lamtau2_fiduc_fidi1} shows the effect each of these
regions has on the observed line profile.

In Fig. \ref{fig:lamtau2_fiduc_fidi1} we show the optical depth as a
function of observed wavelength, subdivided by the velocity of the IGM
gas responsible for the optical depth, for the fiducial galaxy with
the dynamic and simple IGM models.  We label the velocity of the IGM
at radius $r$ from the galaxy $v(r)$; $\vs=222~\kms$ for the dynamic
model, and $\vs=373~\kms$ for the simple model.  When examining the
observed line profile at a particular wavelength $\lambda$, it is
useful to define the velocity in the frame of the galaxy where the
\lya\ resonance corresponds to that wavelength,
\begin{equation}
\val=c\left[1-\frac{\lambda}{(1+z)\lamal}\right].
\end{equation}

The total optical depth has three prominent regions.  At $\val>\vs$,
the optical depth is dominated by Doppler core scattering from neutral
IGM at $v\simeq\val$.  At $\vvir<\val<\vs$ the optical depth is still
due to Doppler core scattering, but the IGM at $v\simeq\val$ is mostly
ionized.  The third region is at $\val<\vvir$, where the optical depth
is due to IGM gas at $v>\vs$ (the same as the first region), but now
the scattering results from the damping wing of the scattering
cross-section.  Figure \ref{fig:lamtau2_fiduc_fidi1} illustrates that
the central part of the observed line profile depends on the
ionization state of the IGM gas in the Stromgren sphere, and the red
wing of the line profile depends on the size of the Stromgren sphere
and the IGM outside of it.

\begin{figure}
  \includegraphics[width=8.4cm]{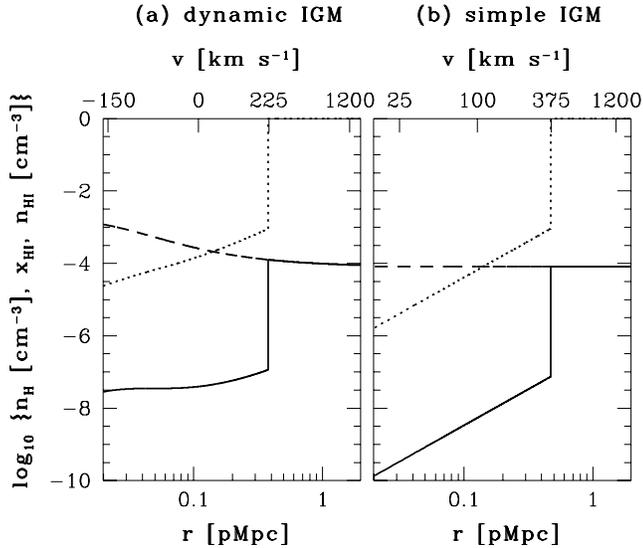}
  \caption{Density and ionization state of the IGM as a function of
  radius and velocity.  (a) Dynamic IGM model.  (b) Simple model.  The
  dashed curves are $\nh$, the total (\hi+\hii) hydrogen number
  density; the dotted curves are $\xhi$, the neutral hydrogen
  fraction; and the solid curves are $\nhi\equiv\xhi\nh$.  Both plots
  are for the fiducial galaxy with radius in physical Mpc and velocity
  in the frame of the galaxy.}
  \label{fig:raddhii2_fiduc_fidi1}
\end{figure}

Figure \ref{fig:raddhii2_fiduc_fidi1} quantitatively compares the IGM
density computed for both the dynamic and simple IGM models.  In the
dynamic model, $\rs=0.38~\pmpc$, versus $\rs=0.47~\pmpc$ for the
simple model.  The larger Stromgren sphere in the simple model is a
consequence of the lower IGM density.

\begin{figure}
  \includegraphics[width=8.4cm]{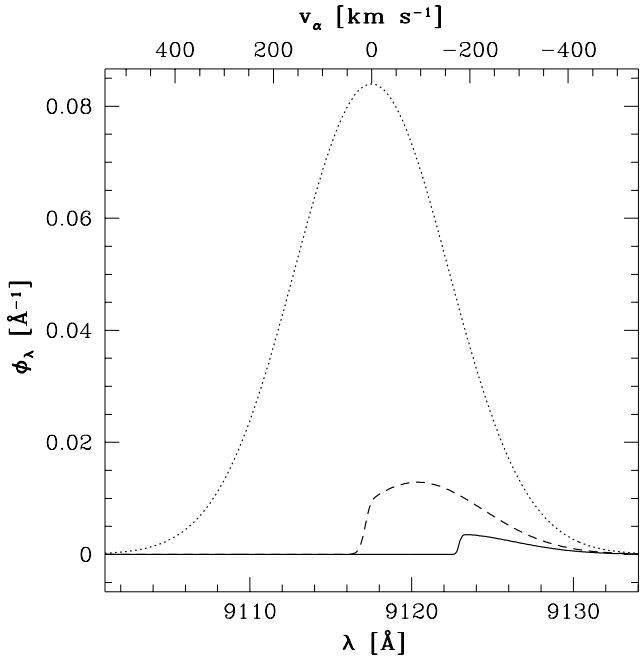}
  \caption{\lya\ line profiles.  The dotted curve is the
  intrinsic \lya\ line profile of the fiducial galaxy.  The solid
  curve is the observed \lya\ line profile using the dynamic IGM
  model.  The dashed curve is the observed line using the simple IGM
  model.  The velocity where the \lya\ resonance corresponds to the
  wavelength on the bottom axis is shown as $\val$, on the top axis.}
  \label{fig:olamline2_fiduc_fidi1}
\end{figure}

Figure \ref{fig:olamline2_fiduc_fidi1} shows the observed \lya\ line
profiles per unit wavelength, $\phil$, for the dynamic (solid curve)
and simple (dashed curve) IGM models, compared to the intrinsic line
profile (dotted curve) of the fiducial galaxy.  The line profiles are
plotted against observed wavelength and also $\val$.

For both the dynamic and simple models, the IGM outside of the virial
radius, though highly ionized out to the Stromgren radius, has
sufficient optical depth to completely suppress the \lya\ line with
resonant scattering from the Doppler core of the absorption
cross-section.  Thus the wavelength range over which the line is
completely suppressed corresponds to $\val\ga\vvir$, where $\vvir$ is
the velocity of the IGM at the virial radius.  For the dynamic model
$\vvir=-160~\kms$, in contrast to $15~\kms$ for the simple model.

Because IGM gas is assumed to ionize completely as it crosses the
virial radius, the line redward of $\vvir$ is not scattered by the
Doppler core of the absorption profile of any IGM gas.  However, the
observed lines are still depressed compared to the intrinsic profile
at $\val<\vvir$ due to scattering by the fully neutral IGM outside the
Stromgren sphere, via the red damping wing of the absorption
cross-section.  In summary, the optical depth at some wavelength (and
thus $\val$) can be thought of as a sum of the optical depth due to
Doppler core scattering by IGM gas at $v=\val$, and optical depth due
to damping wing scattering by IGM gas at $v\geq\vs$.  In
Fig. \ref{fig:olamline2_fiduc_fidi1}, the difference between the
dynamic and simple model predictions at $\val=-100~\kms$ is primarily
a consequence of the presence of neutral gas at $v=-100~\kms$ in the
dynamic model.  At $\val=-300~\kms$, the difference is entirely a
consequence of neutral gas between +225 and $+375~\kms$ (the velocity
range over which the IGM is neutral in the dynamic model but ionized
in the simple model).

The integrated observed line flux for the dynamic model is only 1.6
per cent of the intrinsic line, compared to 11 per cent for the simple
model.  Thus incorporating a dynamic model for the IGM is an important
consideration in connecting theoretical predictions for \lya\ emitters
into observed counts, and for interpreting the ionization state of the
IGM through \lya\ emission line observations.

\subsection{Galaxy/halo parameter dependence}

In the previous section we showed that our dynamic IGM model has a
substantial impact on the observed \lya\ line from a galaxy at high
redshift.  In this section we illustrate how the observed line depends
on properties of the galaxy and its halo.  The parameters we explore
are the halo mass, galaxy star-formation rate, age of star formation,
escape fraction of ionizing radiation, and relation between the
velocity width of the emitted line and the halo circular velocity.
All figures are for the dynamic IGM model only; varying the parameters
would have analogous effects under the simple IGM model.

For each parameter, we show plots analogous to
Fig. \ref{fig:raddhii2_fiduc_fidi1} and
Fig. \ref{fig:olamline2_fiduc_fidi1}, as appropriate.  From these
plots the virial radius and Stromgren radius can be read off, as well
as the corresponding IGM infall velocities.  The line profile figures
show the effect of our model parameters, if any, on the intrinsic line
profile, and the effect on the observed line profile.  

\paragraph*{Halo mass}

\begin{figure}
  \includegraphics[width=8.4cm]{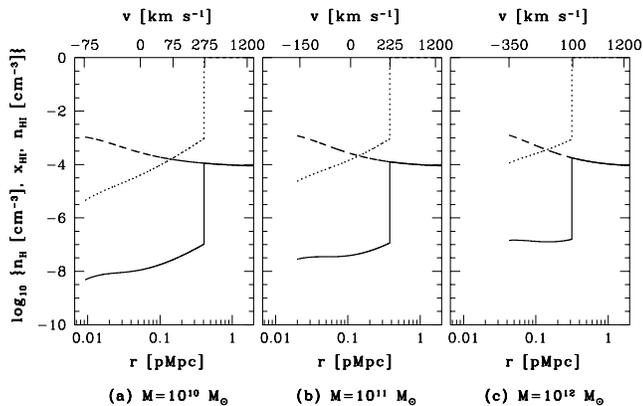}
  \caption{As in Fig. \ref{fig:raddhii2_fiduc_fidi1}, but note
  different radial scale.  The curves are truncated at $\rvir$.  (a)
  Halo mass $M=10^{10}~\msun$.  (b) $M=10^{11}~\msun$ (fiducial
  model).  (c) $M=10^{12}~\msun$.}
  \label{fig:raddhim3_fiduc_fidm10_fidm12}
\end{figure}

In Fig. \ref{fig:raddhim3_fiduc_fidm10_fidm12} we recompute the IGM
properties for the fiducial model, except that we set the halo mass to
$10^{10}$ or $10^{12}~\msun$.  Though more massive halos have larger
values of $\rvir$ (the radii at which the curves terminate), they have
more negative values of $\vvir$.  More massive halos also have
slightly higher IGM densities around them.  The combination of larger
virial radii (ionizing flux falls off as $r^{-2}$) and slightly higher
IGM densities leads to larger neutral fractions within the Stromgren
spheres of more massive halos.  The Stromgren spheres of more massive
halos are slightly smaller (due to larger densities within them), but
$\vs$ decreases more dramatically with increasing halo mass.

\begin{figure}
  \includegraphics[width=8.4cm]{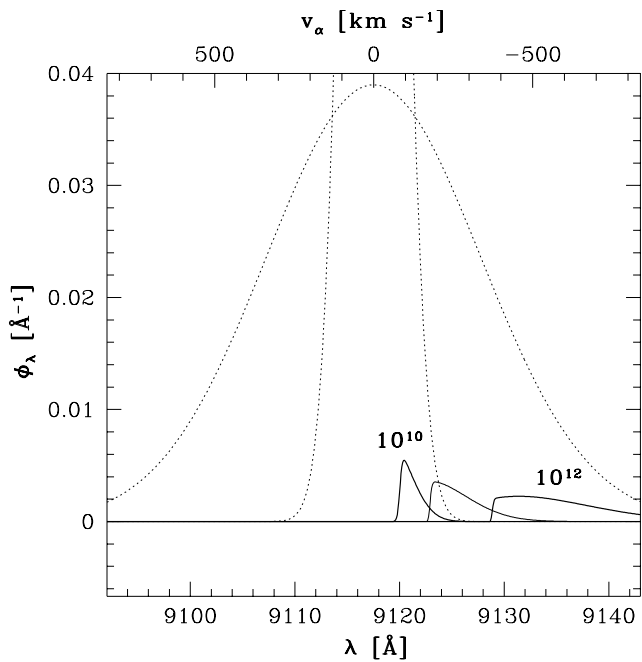}
  \caption{As in Fig. \ref{fig:olamline2_fiduc_fidi1}, but note
  different axis scales.  The labels show $M$ (in $\msun$) for the
  corresponding intrinsic and observed line profiles.  The fiducial
  model $M=10^{11}~\msun$ is the lighter curve, and the intrinsic
  fiducial line profile is not shown.}
  \label{fig:olamlinem3_fiduc_fidm10_fidm12}
\end{figure}

In Fig. \ref{fig:olamlinem3_fiduc_fidm10_fidm12} we show the
corresponding intrinsic and observed \lya\ line profiles.  More
massive halos have broader intrinsic and observed lines (because
$\vd\equiv\fv\vc$ and $\vc$ increases for more massive halos).  A new
prediction of the dynamic IGM model is that the offset between the
intrinsic line center and the blue edge of the observed line (or,
similarly, the centroid of the observed line) increases with
increasing halo mass.  This conclusion is somewhat dependent on the
assumption that the intrinsic \lya\ emission line is centered on the
systemic velocity of the galaxy.  In practice obtaining the intrinsic
\lya\ line shape may be impossible, but a more complicated model of
the intrinsic line may be passed through our IGM model to produce a
predicted observed line profile, which will always have a lower limit
in wavelength corresponding to $\val=\vvir$.

\paragraph*{Star-formation rate}

\begin{figure}
  \includegraphics[width=8.4cm]{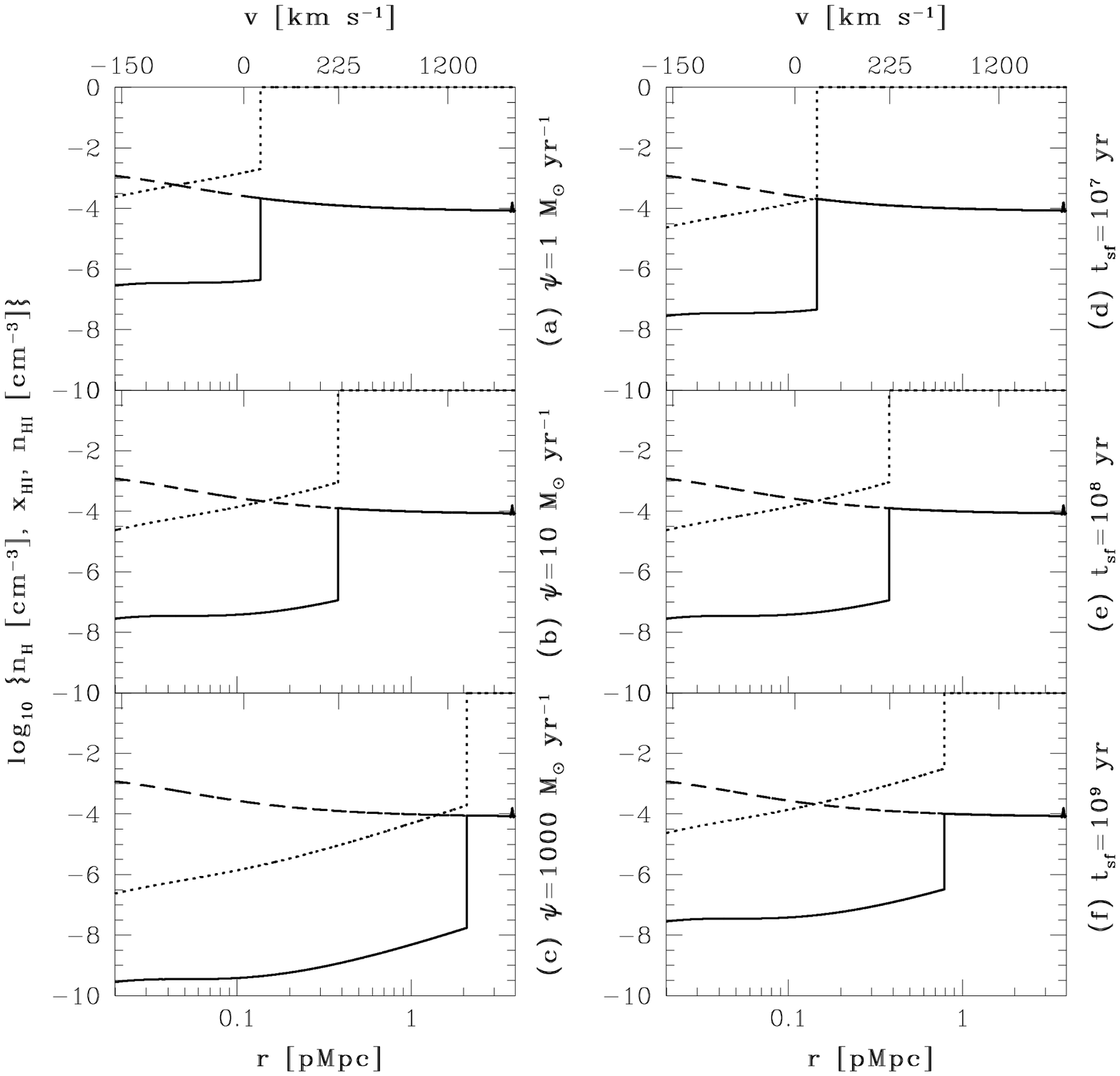}
  \caption{As in Fig. \ref{fig:raddhii2_fiduc_fidi1}.  (a)
  Star-formation rate $\psi=1~\sfr$.  (b) $\psi=10~\sfr$ (fiducial
  model).  (c) $\psi=1000~\sfr$.  See
  Fig.~\ref{fig:raddhier6_fiduc_fide0_fide1_fidr0_fidrm5_fidr2}c for
  $\psi=10~\sfr$.  (d) Age of star formation $\tsf=10^7~\yr$.  (e)
  $\tsf=10^8~\yr$ (fiducial model).  (f) $\tsf=10^9~\yr$.}
  \label{fig:raddhist5_fiduc_fids1_fids1000_fidt7_fidt9}
\end{figure}

\begin{figure}
  \includegraphics[width=8.4cm]{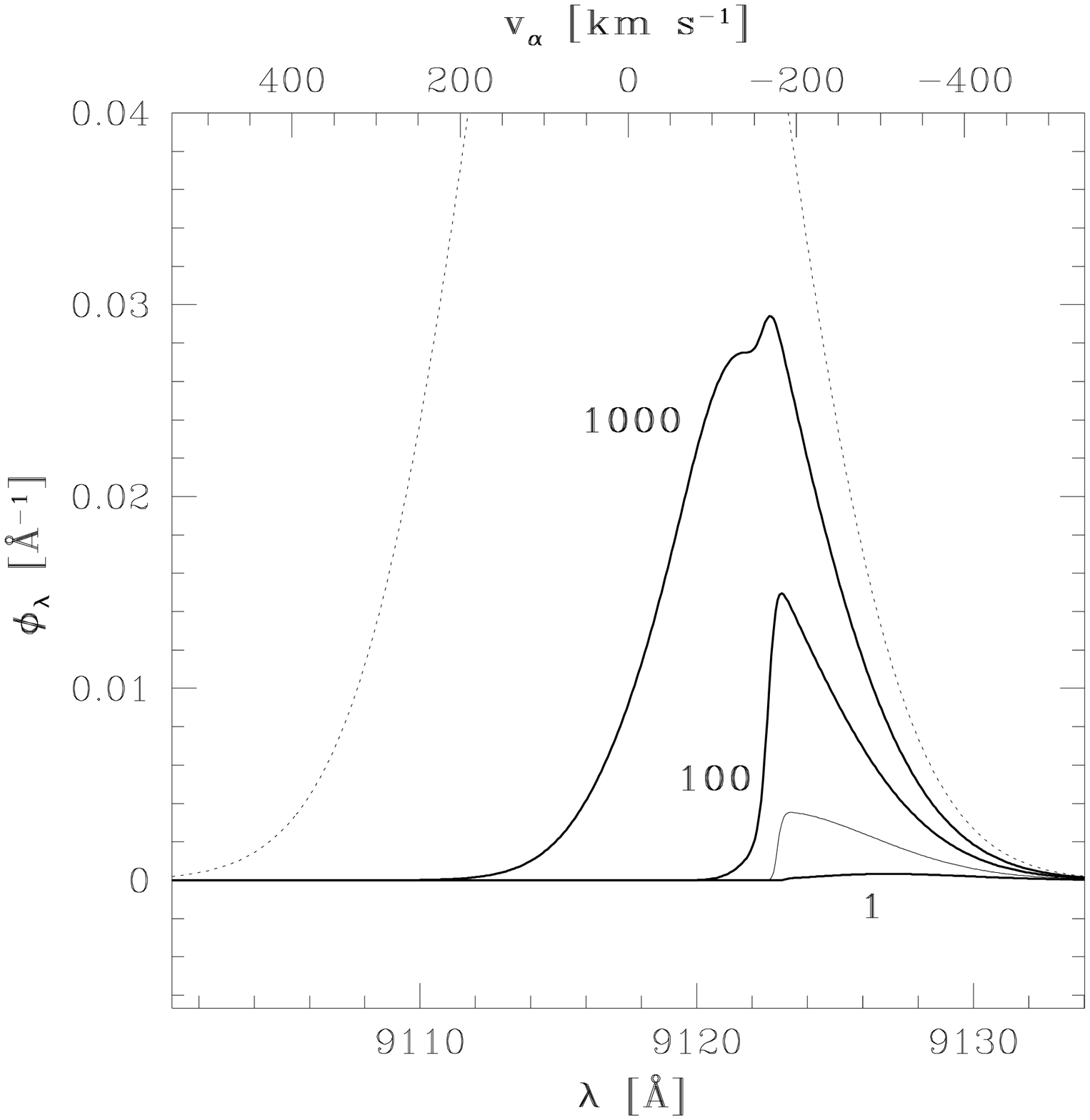}
  \caption{As in Fig. \ref{fig:olamline2_fiduc_fidi1}, but different
  vertical scale.  The labels show star-formation rate (in $\sfr$) for
  the corresponding observed line profiles.  The fiducial model
  $\psi=10~\sfr$ is the lighter curve, and the intrinsic fiducial line
  profile is the dotted curve.}
  \label{fig:olamlines4_fiduc_fids1_fids100_fids1000}
\end{figure}

In Figs. \ref{fig:raddhist5_fiduc_fids1_fids1000_fidt7_fidt9} and
\ref{fig:olamlines4_fiduc_fids1_fids100_fids1000} we recompute the IGM
properties and \lya\ line profiles for the fiducial model, except that
we set the SFR to 1, 100, or 1000~$\sfr$.  Increasing the SFR
increases both the size of the Stromgren sphere and the ionization
fraction inside.  If we decrease the SFR from the fiducial value of
$10~\sfr$ to $1~\sfr$ then the neutral IGM encroaches all the way to
$\vs=25~\kms$ (compared to $\vs=225~\kms$ in the fiducial case), and
thus the damping wing almost completely wipes out the observed line.
Increasing the SFR to $100~\sfr$, the Stromgren sphere expands out to
$\vs=650~\kms$.  Though there is a decrease in the neutral fraction
within the Stromgren sphere, this effect is not sufficient to lift the
complete suppression of the observed line at $\val\ga\vvir$.  However,
the greater distance (in velocity) to the fully neutral IGM means that
the observed line redward of of $\val\simeq\vvir$ is less damped than
in the fiducial case.  A further increase of the SFR to $1000~\sfr$
expands the Stromgren sphere yet further, resulting in very little
damped absorption at $\val\la\vvir$.  Moreover, the neutral fraction
inside part of the Stromgren sphere is now small enough that
scattering by the cross-section Doppler core no longer completely
suppresses the line at $\val>\vvir$.  Note the notch in the spectrum
at $\val=\vvir$: this is the feature predicted in QSO spectra by
\citet{bar03a}.

\paragraph*{Age of star formation}

\begin{figure}
  \includegraphics[width=8.4cm]{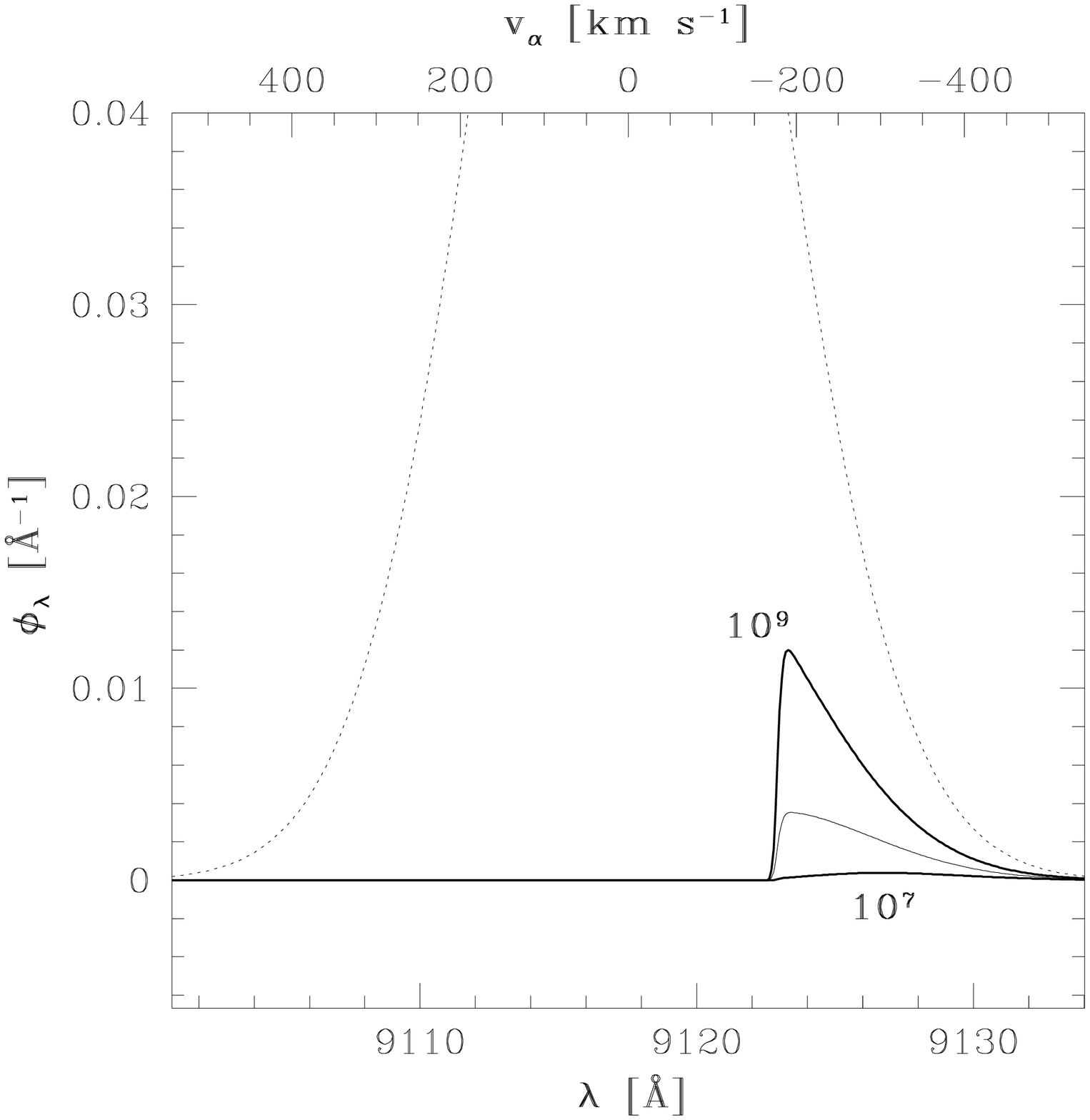}
  \caption{As in Fig. \ref{fig:olamline2_fiduc_fidi1}, but different
  vertical scale.  The labels show the age of star-formation (in
  $\yr$) for the corresponding observed line profiles.  The fiducial
  model $\tsf=10^8~\yr$ is the lighter curve, and the intrinsic
  fiducial line profile is the dotted curve.}
  \label{fig:olamlinet3_fiduc_fidt7_fidt9}
\end{figure}

In Figs. \ref{fig:raddhist5_fiduc_fids1_fids1000_fidt7_fidt9} and
\ref{fig:olamlinet3_fiduc_fidt7_fidt9} we recompute the IGM and line
profile properties for the fiducial model, except that we set the age
of star formation to $10^7$ or $10^9~\yr$.  Changing $\tsf$ changes
the size of the Stromgren sphere, but does not change the ionization
balance within it.  Thus the effect on the observed line profile is
only to change the amount of damping of the line at $\val<\vvir$.

\paragraph*{Escape fraction of ionizing photons}

\begin{figure}
  \includegraphics[width=8.4cm]{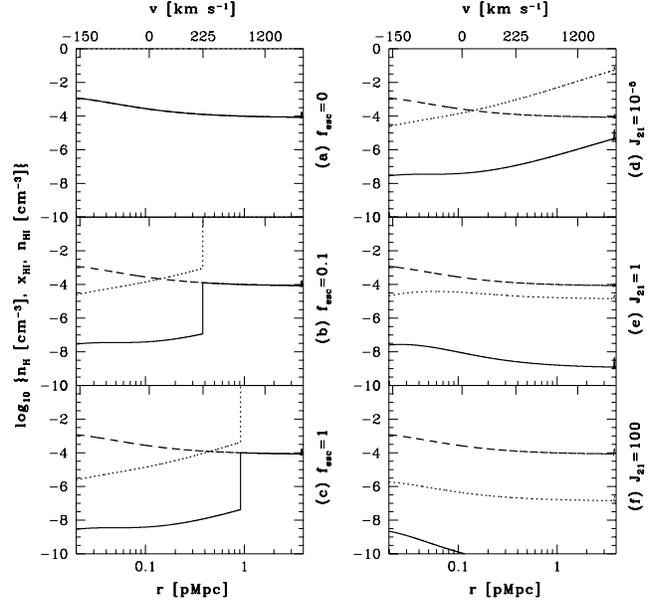}
  \caption{As in Fig. \ref{fig:raddhii2_fiduc_fidi1}.  (a) Ionizing
  photon escape fraction $\fesc=0$.  (b) $\fesc=0.1$ (fiducial model).
  (c) $\fesc=1$.  (d) Ionizing background $\jto=10^{-5}$.  (e)
  $\jto=1$.  (f) $\jto=100$.}
  \label{fig:raddhier6_fiduc_fide0_fide1_fidr0_fidrm5_fidr2}
\end{figure}

In Fig.~\ref{fig:raddhier6_fiduc_fide0_fide1_fidr0_fidrm5_fidr2} we
recompute the IGM properties for the fiducial model, except that we
set $\fesc=0$ or 1.  Scaling $\fesc$ has the same effect as scaling
the star-formation rate by the same factor, so the fiducial model
modified to $\fesc=1$ is the same as the fiducial model modified to
$\psi=100~\sfr$ (see above).  In the case of vanishing escape
fraction, for any reasonable value of the SFR the line will
unobservable, because a neutral IGM down to $\vvir$ allows only
$4\times10^{-5}$ of the intrinsic line to be observed.  See also
Section~\ref{sec:obslae}.

\paragraph*{Velocity width factor}

\begin{figure}
  \includegraphics[width=8.4cm]{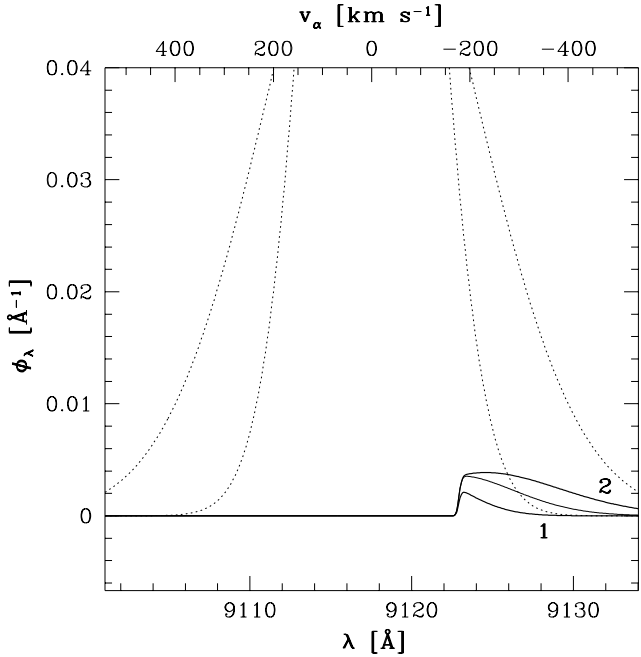}
  \caption{As in Fig. \ref{fig:olamline2_fiduc_fidi1}, but different
  vertical scale.  The labels show the conversion factor $\fv$ between
  Doppler line width and halo circular velocity for the corresponding
  intrinsic and observed line profiles.  The fiducial model $\fv=1.5$
  is the lighter curve, and the intrinsic fiducial line profile is not
  shown.}
  \label{fig:olamlinev3_fiduc_fidv1_fidv2}
\end{figure}

In Fig. \ref{fig:olamlinev3_fiduc_fidv1_fidv2} we recompute line
profile properties for the fiducial model, except that we set $\fv=1$
or 2 ($\fv$ does not influence the IGM properties).  Larger values of
$\fv$ broaden the line, resulting in a greater observable fraction of
the intrinsic line.  In the limit of much larger values of $\fv$, more
of the line would be transmitted, but the line width would be so large
that spectroscopic detection of the line would become more difficult
than for a narrower line with less flux.  However, for the values of
$\fv$ we consider, which should bound the actual values, the observed
line is still relatively narrow and larger values of $\fv$ produce
more easily observed lines.

\subsection{IGM ionization dependence}
\label{sec:igmiondep}

In this section we vary the ionization state of the IGM to show how
this influences the observed \lya\ line.  First we will assume, in
contrast to the IGM models considered so far, that there is a mean
ionizing background in the IGM that contributes to the ionization of
the IGM around the halo, in addition to the contribution from the
galaxy itself.  We assume that the universe was reionized before star
formation turned on in the halo (that is, reionization occurred longer
than $\tsf$ ago), and since then the background has been fixed at
\begin{equation}
\jto \equiv \frac{J_\nu}{10^{-21}~\mathrm{erg\, s^{-1}\, cm^{-2}\,
Hz^{-1}\, sr^{-1}}}.
\end{equation}
For the fiducial values of $z=6.5$ and $\tsf=10^8~\yr$, that
corresponds to reionization before $z=7.2$.

\begin{figure}
  \includegraphics[width=8.4cm]{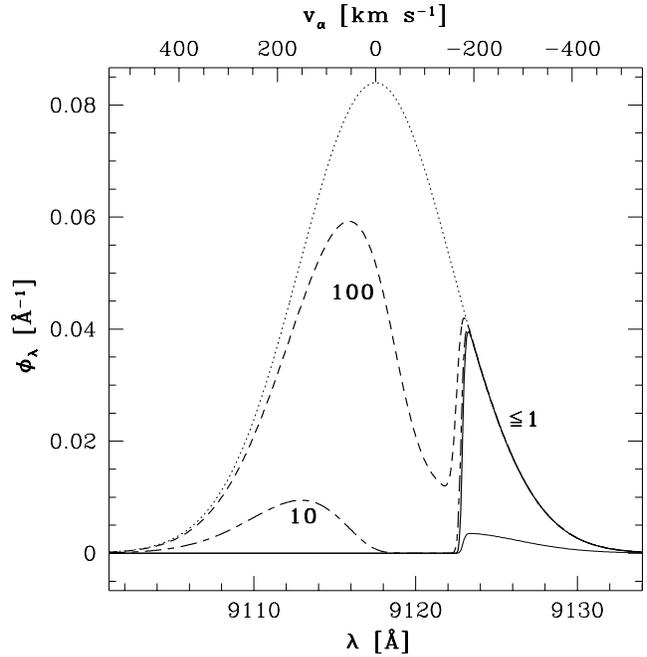}
  \caption{As in Fig. \ref{fig:olamline2_fiduc_fidi1}.  The labels
  show ionizing background strength $\jto$ for the corresponding
  observed line profiles.  The fiducial model with no ionizing
  background is the lighter curve, and the intrinsic fiducial line
  profile is the dotted curve.}
  \label{fig:olamliner4_fiduc_fidr0_fidr1_fidr2}
\end{figure}

In Figs. \ref{fig:raddhier6_fiduc_fide0_fide1_fidr0_fidrm5_fidr2} and
\ref{fig:olamliner4_fiduc_fidr0_fidr1_fidr2} we show the properties of
the IGM and \lya\ line profiles for values of the ionizing background
that bracket the physically interesting regimes.  First we discuss
$\jto=10^{-5}$, a value so small that the IGM ionization state inside
the galaxy's light sphere is completely dominated by the galaxy's
ionizing radiation.  The light sphere has a radius of
$c\,\tsf=31~\pmpc$ for $\tsf=10^8~\yr$, corresponding to an IGM
velocity of almost $2.5\times10^4~\kms$.  Consequently, the IGM
ionization state outside of the light sphere is irrelevant.

Even though the ionizing background does not contribute at all to the
ionization rate (see eq. \ref{eq:ionrate}) at any radius of interest,
the pre-reionization of the universe before the galaxy turned on
allows its Stromgren sphere to expand at near light speed.
Consequently even the IGM at the velocity of $1200~\kms$ has a neutral
fraction of 0.01.  By the Gunn-Peterson trough test, this IGM wouldn't
even qualify as reionized.  However, this level of ionization
completely eliminates any damping wing optical depth.  Thus the
observed line profile follows the intrinsic line profile from the red
wing up to $\val=\vvir$, then goes to zero as in the fiducial case.
This is the observed behavior for any ionizing background with
$\jto\la1$, the range observed at $z<5$ \citep{sco00}.

Ionizing backgrounds in the range $\jto\sim10$--100 strongly ionize
the mean density IGM.  Since the IGM density decreases from a local
maximum at $\rvir$ toward the cosmic mean as a function of increasing
radius, IGM gas just outside of the virial radius still strongly
scatters the intrinsic line at $\val$ corresponding to the velocity of
the overdense gas.  As the density decreases, though, the neutral
fraction drops as well, and the combination leads to greater
transmission of the intrinsic line \textit{toward the blue side of the
line}.  This would generate an unusually-shaped observed line with two
peaks, one of which lies to the blue of the intrinsic line center.

At yet higher values of $\jto\ga10^3$, the IGM is so strongly ionized
that the entire intrinsic line would be observed.  We note again that
we expect $\jto\la1$, though we are unaware of any predictions for
$\jto$ at $z>6$ in models that reproduce the \textit{WMAP} optical
depth value.

\begin{figure}
  \includegraphics[width=8.4cm]{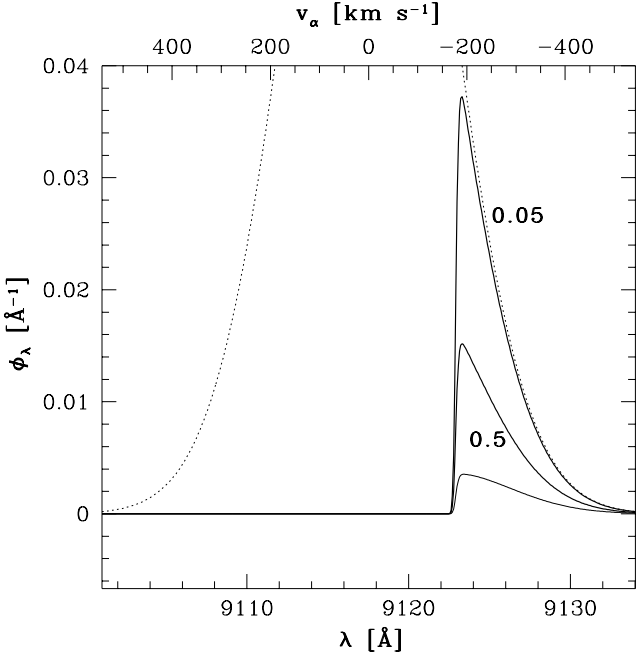}
  \caption{As in Fig. \ref{fig:olamline2_fiduc_fidi1}, but with
  different vertical scale.  The labels show the neutral fraction of
  the IGM outside of the galaxy's Stromgren sphere for the
  corresponding observed line profiles.  The fiducial model with a
  neutral IGM is the lighter curve, and the intrinsic fiducial line
  profile is the dotted curve.}
  \label{fig:olamlinex3_fiduc_fidx0.5_fidx0.05}
\end{figure}

Alternately, rather than considering a fixed ionizing background in a
reionized universe, we can fix the neutral fraction of the IGM outside
of the galaxy's Stromgren sphere, $\xigm$.  This approach is more
useful for considering a weakly-ionized IGM with $\xigm\sim0.5$
\citep[e.g.,][where a fully reionized universe almost entirely
recombines]{cen03a}.
Figure~\ref{fig:olamlinex3_fiduc_fidx0.5_fidx0.05} shows the observed
\lya\ line profiles of the fiducial galaxy when $\xigm$ is set to 0.5
and 0.05.  As expected, the decreased \hi\ abundance in the `neutral'
IGM means that the red wing of the \lya\ line is less suppressed by
damping wing scattering.  There is no effect on the Doppler scattering
that scatters the center of the intrinsic line.  See also
Section~\ref{sec:obslae}.

\subsection{Redshift dependence}

Last, we turn to the effects due solely to changes in redshift.  In
the future there will (hopefully) be surveys for strong \lya\ emission
lines at several different redshifts at $z>6$.  We study redshift
dependence by keeping the other free parameters fixed; realistically,
other parameters, such as star-formation rate, age of star formation,
escape fraction, and the typical halo mass, may also depend on
redshift.  Comparison of the galaxy/halo parameter dependencies
described above in conjunction with the effects shown here can be used
to predict trends for a specific model of the evolution of the
galaxy/halo parameters.  Disentangling the redshift dependence of the
IGM ionization state from the other redshift dependences may still be
very difficult in practice.

\begin{figure}
  \includegraphics[width=8.4cm]{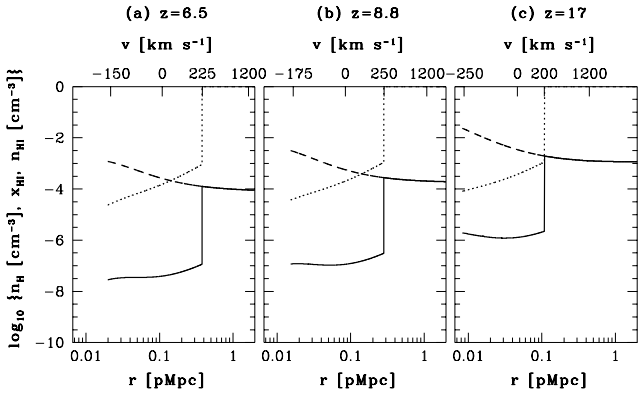}
  \caption{As in Fig. \ref{fig:raddhii2_fiduc_fidi1}, but note
  different radial and velocity scales.  The curves are truncated at
  $\rvir$.  (a) Redshift $z=6.5$ (fiducial model).  (b) $z=8.8$.  (c)
  $z=17$.}
  \label{fig:raddhiz3_fiduc_fidz8.8_fidz17}
\end{figure}

\begin{figure}
  \includegraphics[width=8.4cm]{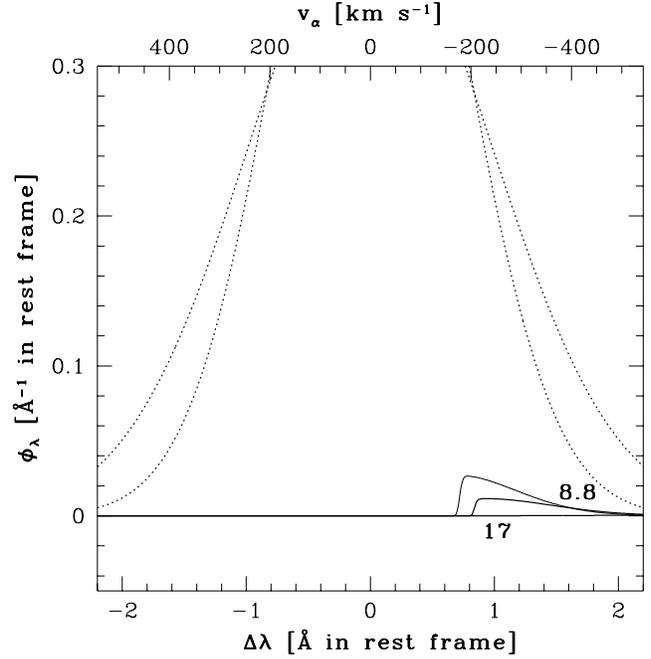}
  \caption{\lya\ line profiles.  The wavelength scale is expressed as
  an offset from the \lya\ resonance in the frame of the galaxy,
  chosen so that the $\val$ scale is the same as in
  Fig. \ref{fig:olamline2_fiduc_fidi1}.  The labels show the redshift
  for the corresponding observed line profiles.  The narrower
  intrinsic profile (dotted curve) is for $z=8.8$; the wider intrinsic
  profile (dotted curve) is for $z=17$.  The fiducial model $z=6.5$ is
  the lighter curve, and the intrinsic fiducial line profile is not
  shown.}
  \label{fig:dlamlinez3_fiduc_fidz8.8_fidz17}
\end{figure}

Proceeding with simple cases to understand the expected effects, we
place the fiducial halo at $z=8.8$ and $z=17$.  Figures
\ref{fig:raddhiz3_fiduc_fidz8.8_fidz17} and
\ref{fig:dlamlinez3_fiduc_fidz8.8_fidz17} show the IGM and \lya\ line
profile properties.  Because we keep the mass of the halo fixed, the
virial radius shrinks at increasing redshift.  This results in an
increasingly negative velocity of the IGM at the virial radius.  The
IGM is also denser at higher redshift, resulting in higher neutral
fractions when the ionization balance is calculated.  An additional
effect is the increased circular velocity of the halo, which broadens
the intrinsic line.

\section{Galactic winds}
\label{sec:galwind}

We now consider another, possibly crucial, alteration to the model of
\lya\ emitter plus IGM scattering described so far.  High velocity
($\sim360~\kms$) redshifts of \lya\ emission compared to other
interstellar features are observed in $z=3$ Lyman Break Galaxies
\citep{sha03}; QSO absorption line cross-correlation suggests that
these winds may have blown to large distances
\cite[$\sim0.2$~pMpc,][]{ade03}.

Galactic winds, if present in galaxies at $z>6$, may have two
important consequences for the interpretations of \lya\ emission from
galaxies during reionization.  The first is that as the wind blows
through the IGM, it may collisionally ionize the hydrogen it passes,
increasing the ionization fraction of this gas \citep{ade03}.  The
second effect is that the intrinsic \lya\ line may not have a Doppler
shape centered on the galaxy's redshift.  Instead, it may already
display an asymmetric profile centered considerably to the red of the
systemic redshift \citep[where they find a mean redshift of
$360~\kms$]{sha03}.  Inspection of Fig.~\ref{fig:lamtau2_fiduc_fidi1}
shows that the IGM optical depth for the fiducial galaxy is only
$\sim1$ at redshifts of more than $200~\kms$\footnote{Note that in
observations velocity (and redshift) is measured in the observer's
frame, whereas in our plots $\val$ is in the frame of the emitting
galaxy, hence a line redshifted by $360~\kms$ falls at
$\val=-360~\kms$.}.

\begin{figure}
  \includegraphics[width=8.4cm]{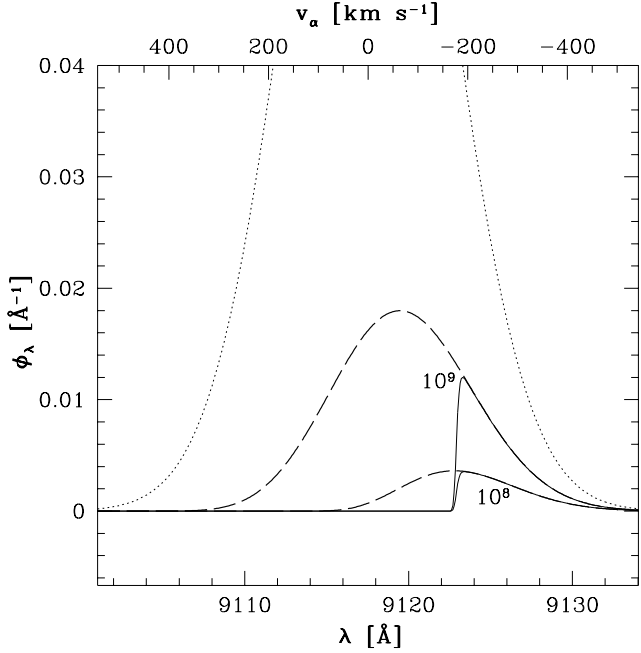}
  \caption{As in Fig. \ref{fig:olamline2_fiduc_fidi1}, but different
  vertical scale.  The labels show star-formation timescale (in $\yr$)
  for the corresponding observed line profiles.  For each $\tsf$, the
  light solid curve is the profile without assuming any winds (as in
  Fig.~\ref{fig:olamlinet3_fiduc_fidt7_fidt9}), and the long-dashed
  curve is the observed line profile assuming a wind blows to the
  Stromgren radius.  The intrinsic fiducial line profile is the dotted
  curve.}
  \label{fig:olamwlinet2_fiduc_fidt9}
\end{figure}

Since we lack a good physical model to calculate wind properties, we
adopt a very simple assumption to illustrate the possible effects of
large-scale winds on the IGM (the first effect described above).  We
assume that a wind blows out to the Stromgren radius of the galaxy
(see Fig.~\ref{fig:raddhist5_fiduc_fids1_fids1000_fidt7_fidt9}),
completely ionizing all of the gas within that radius.
Figure~\ref{fig:olamwlinet2_fiduc_fidt9} shows the observed \lya\ line
profiles including such winds.  The result for the fiducial model is
that the observed line becomes roughly symmetric, though still
centered to the red of the intrinsic line center, and the observed
flux doubles.  If $\tsf$ is changed to $10^9~\yr$, the Stromgren
sphere is much larger (see
Fig.~\ref{fig:raddhist5_fiduc_fids1_fids1000_fidt7_fidt9}) and the
observed flux increases by almost a factor of 5.

The required wind velocity can be calculated by dividing the Stromgren
radii (illustrated in previous plots) by $\tsf$.  For the fiducial
halo, the Stromgren radius is 0.38~pMpc, giving a wind velocity over
$10^8~\yr$ of $3700~\kms$.  Almost any wind model would predict a
smaller wind velocity than this, so though
Fig.~\ref{fig:olamwlinet2_fiduc_fidt9} shows the qualitative effect,
even the $10^8~\yr$ curve is an overestimate of the true effect,
assuming galactic winds blow to $0.2~\pmpc$, as in \citet{ade03}.

Winds don't affect the damping wing suppression of the line from
absorption by neutral (in the case of a non-reionized IGM) gas outside
of the Stromgren sphere (the solid and long-dashed curves match at
$\val<\vvir$).  Thus the depression of the observed line profiles
compared to the intrinsic line profiles of plots like
Fig.~\ref{fig:olamline2_fiduc_fidi1} can be used to estimate the
transmission of a highly-redshifted \lya\ line.  Since the damping
wing optical depth is set by the IGM outside of the Stromgren sphere,
parameters that affect the size of the Stromgren sphere (e.g., $\tsf$,
Fig.~\ref{fig:raddhist5_fiduc_fids1_fids1000_fidt7_fidt9}), and the
ionization state of the IGM (e.g., $\jto$,
Fig.~\ref{fig:raddhier6_fiduc_fide0_fide1_fidr0_fidrm5_fidr2}, and
$\xigm$), have a large influence on the transmission at $\val<\vvir$.
In the next section we illustrate the effect galactic winds may have
on the observed \lya\ line for all of the parameter variations
discussed in Section~\ref{sec:lyares}.

\section{Discussion}
\label{sec:disc}

\subsection{\lya\ line fluxes}
\label{sec:lyaflux}

\begin{figure}
  \includegraphics[width=8.4cm]{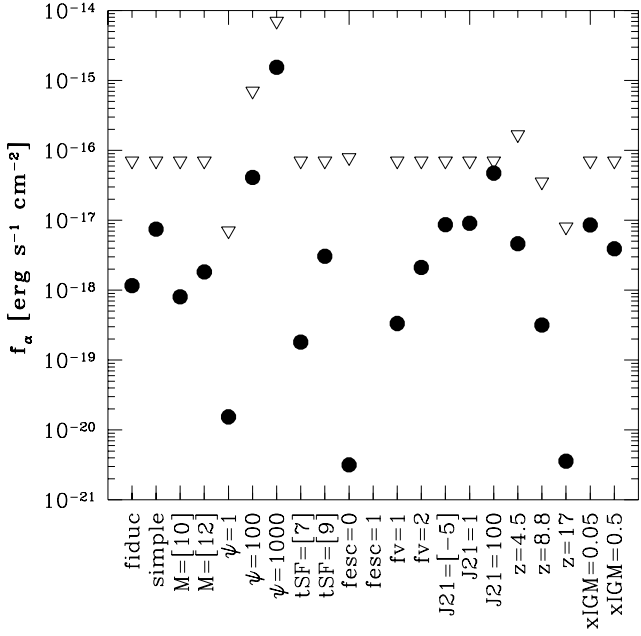}
  \caption{Intrinsic and observed \lya\ line flux as a function of
  model parameters.  Each model plotted is described by a label on the
  bottom axis that refers to the fiducial model in the case of
  `fiduc,' and refers to the parameter varied from the fiducial model
  in the rest of the cases (see Section~\ref{sec:lyares}).  Numbers in
  square braces are the $\log_{10}$ of the parameter value.  The open
  triangles are the intrinsic \lya\ line flux, and the solid circles
  are the observed \lya\ line flux, based on the line profiles
  calculated in Section~\ref{sec:lyares}.}
  \label{fig:paramfal_dfiles}
\end{figure}

First we present three plots to summarize the results of the previous
sections.  Figure~\ref{fig:paramfal_dfiles} shows the intrinsic \lya\
line flux (calculated as described in Section~\ref{sec:inline}) as
open triangles for each model listed on the bottom axis.  The solid
circles show the integrated flux of the observed \lya\ line calculated
from the intrinsic \lya\ line flux and observed line profiles
presented in Section~\ref{sec:lyares}.

\begin{figure}
  \includegraphics[width=8.4cm]{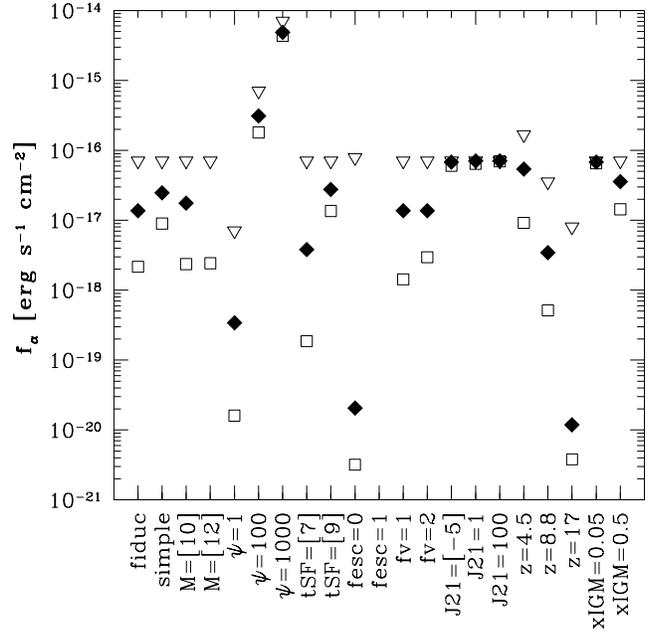}
  \caption{Intrinsic and observed \lya\ line flux as a function of
  model parameters, including possible effects due to galactic winds.
  The model labelling is described in Fig.~\ref{fig:paramfal_dfiles}.
  The open triangles are the intrinsic \lya\ line flux, as in
  Fig.~\ref{fig:paramfal_dfiles}.  The open squares are the integrated
  flux of the observed \lya\ line calculated assuming a galactic wind
  completely ionizes the IGM out to the Stromgren radius of the
  galaxy.  The solid diamonds are the integrated flux of the observed
  \lya\ line calculated assuming the intrinsic \lya\ line is emitted
  entirely at $360~\kms$ redward of the systemic \lya\ wavelength.
  See Section~\ref{sec:galwind}.}
  \label{fig:paramwfal_dfiles}
\end{figure}

\begin{figure}
  \includegraphics[width=8.4cm]{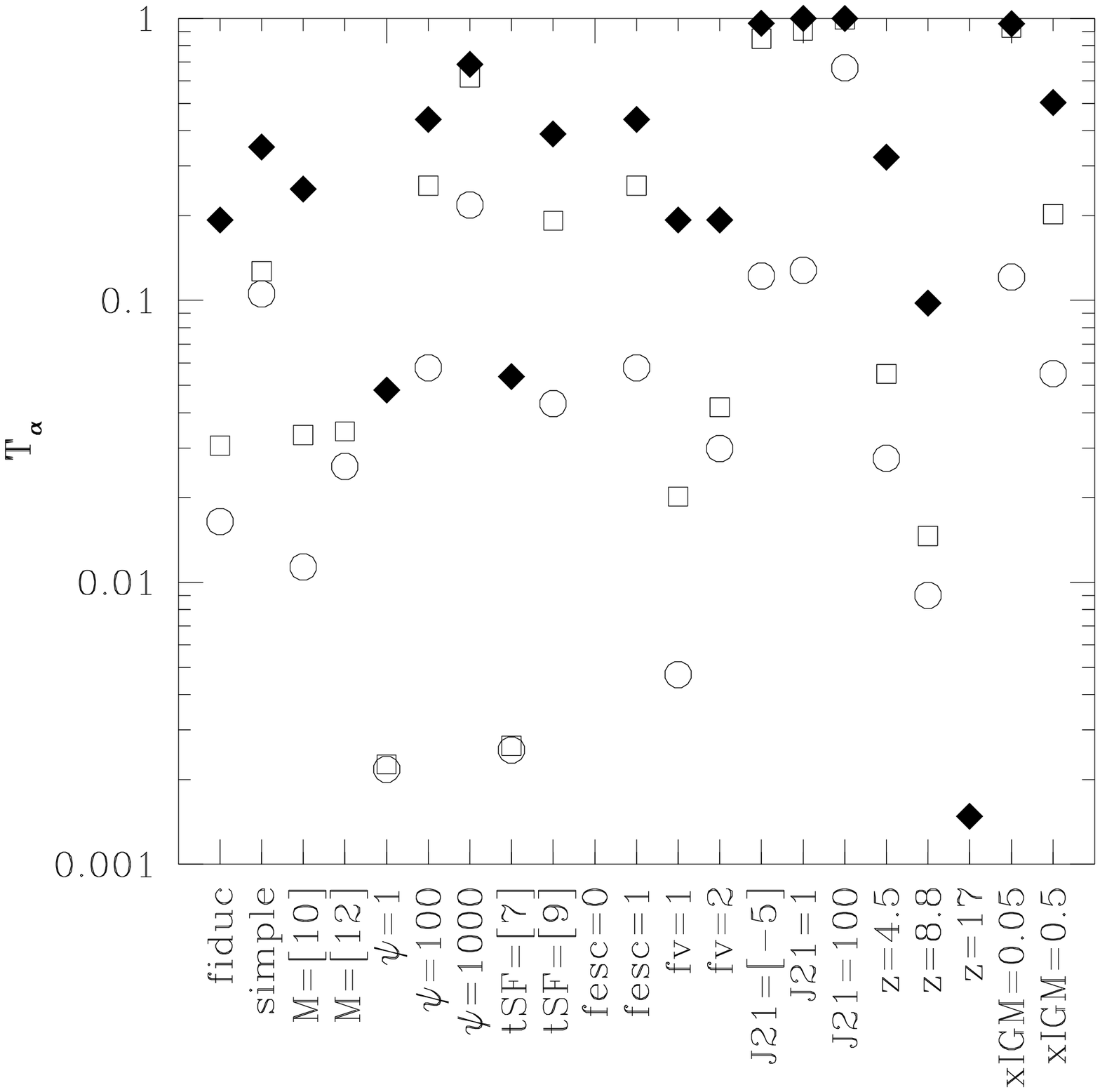}
  \caption{Ratio of observed to intrinsic \lya\ line flux as a
  function of model parameters, for both no wind and wind models.  The
  model labelling is described in Fig.~\ref{fig:paramfal_dfiles}.  The
  point shapes are the same as in Figs.~\ref{fig:paramfal_dfiles} and
  \ref{fig:paramwfal_dfiles}: open circles show no-wind models, open
  squares show models where a strong wind clears the Stromgren sphere
  of IGM, and solid diamonds show models where the intrinsic \lya\ is
  concentrated at $360~\kms$ redward of the systemic redshift.}
  \label{fig:paramt_dfiles}
\end{figure}

Figure~\ref{fig:paramwfal_dfiles} shows the intrinsic \lya\ line flux
as open triangles for each model, as in
Fig.~\ref{fig:paramfal_dfiles}.  Here the open squares are the
integrated flux of the observed \lya\ line calculated assuming a
galactic wind completely ionizes the IGM out to the Stromgren radius
of the galaxy (see Section~\ref{sec:galwind}; this is probably an
overestimate of the expected effect, but comparison with
Fig.~\ref{fig:paramfal_dfiles} shows that for many models this is a
small effect regardless).  The solid diamonds are the integrated flux
of the observed \lya\ line calculated assuming the intrinsic \lya\
line is emitted entirely at $360~\kms$ redward of the systemic \lya\
wavelength (see Section~\ref{sec:galwind}).  The consequence of moving
the intrinsic line to the red is that the observed line flux is
generally at least 20 per cent of the intrinsic line flux for most of
the models (independent of whether of not the wind blows into the
IGM).  This is illustrated more explicitly in
Figure~\ref{fig:paramt_dfiles}.  However, more massive galaxies with
$\vvir\la360~\kms$ (such as our $M=10^{12}~\msun$ model) would require
\lya\ lines even more redshifted than $360~\kms$ to make a difference
compared to the no-wind model.  There is no solid diamond for
$M=10^{12}~\msun$ in either Fig.~\ref{fig:paramwfal_dfiles} or
Fig.~\ref{fig:paramt_dfiles} because the IGM transmission at
$\val=-360~\kms$ is formally zero.  A realistic line profile (not
concentrated entirely at $\val=-360~\kms$) would result in a flux
similar to the no-wind model.

Figure~\ref{fig:paramt_dfiles} shows the transmission of the \lya\
line, $T_\alpha$, defined as the ratio of the observed to intrinsic
\lya\ line flux.  The shapes of the symbols match their shapes in
Figs.~\ref{fig:paramfal_dfiles} and \ref{fig:paramwfal_dfiles}.  Quite
generally, an intrinsic \lya\ line redshifted by $360~\kms$ produces
an observed line suppressed by less than a factor of 5, for galaxies
at $z=6.5$ in a neutral IGM.  If the IGM is reasonably ionized, a
redshifted \lya\ line may survive almost unscattered (e.g., solid
diamond for $\xigm=0.05$ model), whereas if the intrinsic line is at
the systemic redshift in those models, the observed line is almost an
order of magnitude weaker (e.g., open circle for $\xigm=0.05$ model).

\subsection{Observed $z=6.5$ \lya\ emitters}
\label{sec:obslae}

To compare observed \lya\ line fluxes to our predictions, we need an
estimate of the intrinsic line flux.  The observed \lya\ line flux
provides a lower limit to the intrinsic flux, but current observations
of $z=6.5$ do provide one avenue to estimate the intrinsic flux.  The
UV continuum to the red of the \lya\ is a star-formation rate
indicator \citep{ken98}; near-IR photometry of $z=6.5$ galaxies
provides an estimate of their rest-frame UV continua.  Two of the
three confirmed $z=6.5$ galaxies have SFRs estimated in this way:
$10~\sfr$ for HCM 6A \citep[hereafter `H,'][]{hu02b}, and $36~\sfr$
for SDF J132415.7+273058 \citep[hereafter `K,'][]{kod03} (in our
cosmology).  As discussed in Section~\ref{sec:inline}, converting the
SFR to an intrinsic \lya\ line flux depends on many parameters, but
for our fiducial choices we estimate intrinsic \lya\ line luminosities
of $3.4\times10^{43}~\lum$ for H and $1.2\times10^{44}~\lum$ for K.
Note that both \citet{hu02a} and \citet{kod03} estimate \lya\ lower
intrinsic luminosities by using an empirical relation between
H$\alpha$ luminosity and SFR \citep{ken98} and converting H$\alpha$
luminosity to \lya\ luminosity.  The corresponding intrinsic \lya\
line fluxes are $6.9\times10^{-17}~\flux$ for H and
$2.4\times10^{-16}~\flux$ for K.
Comparison with the observed \lya\ fluxes (corrected for lensing in
the case of `H') gives estimated observed to intrinsic ratios of
$T_\alpha=0.09$ for both H and K.

From Fig.~\ref{fig:paramt_dfiles} we see that appropriate models with
a neutral IGM and some velocity offset between the intrinsic \lya\
line and the systemic galaxy redshift can produce values of
$T_\alpha\sim0.1$.  Alternately models with a partially or mostly
ionized IGM without any wind effect also have $T_\alpha\sim0.1$.

The redshift of the intrinsic \lya\ line with respect to the systemic
redshift of the galaxy is a clearly key parameter for interpreting
observed \lya\ lines.  Fortunately it may be estimated from
non-resonant recombination lines, such as H$\alpha$ or H$\beta$.  For
galaxies at $z\ga6.5$ these lines fall at $\lambda\ga3.6~\mu$m,
unobservable from the ground in the near-future.  The \textit{James
Webb Space Telescope} (\textit{JWST}) is projected to have sufficient
sensitivity to detect H$\alpha$ and H$\beta$ from the $z=6.5$ galaxies
discovered so far.  Measurements of H$\alpha$ and H$\beta$ will also
provide another SFR estimate (with $\fesc$ no longer entering into the
conversion between SFR and intrinsic \lya\ line flux), as well as a
dust extinction estimate.

An example of an optimistic future observation would be a measurement
of H$\alpha$ and H$\beta$ line redshifts and fluxes for a $z=6.5$
galaxy with a measured \lya\ line and UV continuum.  If the ratio of
H$\alpha$ to H$\beta$ line flux matches the prediction for no
extinction, and the SFR deduced from from those Balmer lines agrees
with the SFR estimated from the UV continuum, then a reasonable
conclusion would be that dust extinction is not very important in the
galaxy\footnote{This conclusion could be strengthened by a stringent
upper limit on, e.g., \heii~1640~\AA, a line that would be present for
an IMF weighted toward very massive stars, \citep{sch03}.}.
Proceeding under the assumption of a normal IMF and no extinction, the
intrinsic \lya\ line flux may then be estimated.  If the \lya\ line
were found to be centered near the Balmer line redshift (assumed to be
the systemic redshift of the galaxy), then we could compare the
measured value of $T_\alpha$ with the circles in
Fig.~\ref{fig:paramt_dfiles}; at that point a $T_\alpha$ could be
converted into an estimate of the neutral fraction of the IGM, though
with plenty of uncertainty associated with unknown parameters such as
$M$, $\tsf$, and $\fv$.  A more clear-cut conclusion may be reached if
the \lya\ line is found to be substantially redshifted compared to the
Balmer redshift and the measured $T_\alpha$ is close to 1.  This would
be a strong signal that the IGM was mostly ionized at that redshift
(see below), and may moreover place an interesting limit on the mass
of the halo, depending on the shape of the observed \lya\ line.

In advance of \textit{JWST}, Fig.~\ref{fig:paramt_dfiles} makes it
clear that drawing any firm conclusions regarding the ionization
fraction of the IGM based only on the observed \lya\ lines and UV
continua of $z=6.5$ is very difficult.

\begin{figure}
  \includegraphics[width=8.4cm]{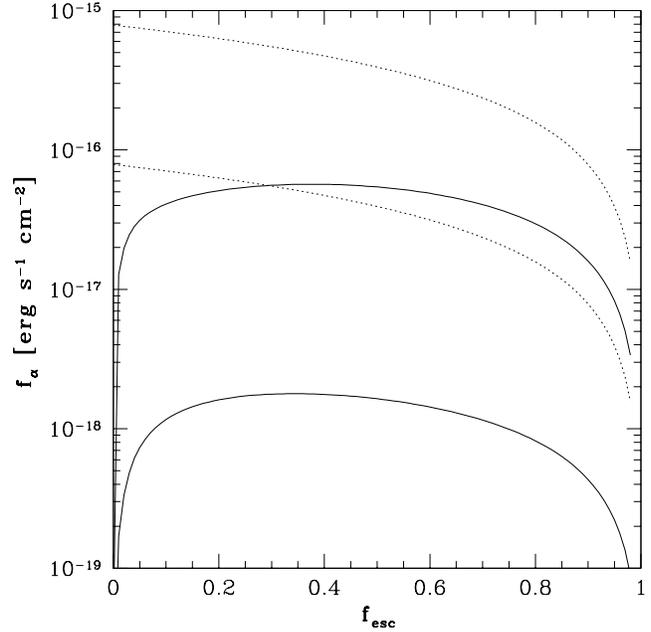}
  \caption{Intrinsic and observed \lya\ line flux as a function of
  ionizing photon escape fraction.  The dotted curves are the
  intrinsic \lya\ line flux; the solid curves are the observed line
  fluxes.  The lower curves are for the fiducial model with $\fesc$
  varied; the upper curves are for the fiducial model modified to
  $\psi=100~\sfr$.}
  \label{fig:lyaplot_lya01.txt_lyas100.txt}
\end{figure}

Before moving on, we present two more plots of interest regarding the
observability of \lya\ lines.  We have seen that large Stromgren
spheres around galaxies allow more of the intrinsic \lya\ line to be
observed (e.g., Fig.~\ref{fig:olamlinet3_fiduc_fidt7_fidt9}).  The
ionization of the IGM is directly proportional to the parameter
$\fesc$, so the largest possible Stromgren sphere is created when
$\fesc=1$.  However, since the intrinsic \lya\ line flux is
proportional to $(1-\fesc)$, the strongest intrinsic \lya\ line is
produced when $\fesc=0$.  Clearly the strongest \textit{observed} line
comes from some intermediate value of $\fesc$.  In
Fig.~\ref{fig:lyaplot_lya01.txt_lyas100.txt} we plot the intrinsic and
observed \lya\ line flux for two models, the fiducial model, and the
$\psi=100~\sfr$ model.  The observed line flux in both models peaks
near $\fesc=0.35$.

\begin{figure}
  \includegraphics[width=8.4cm]{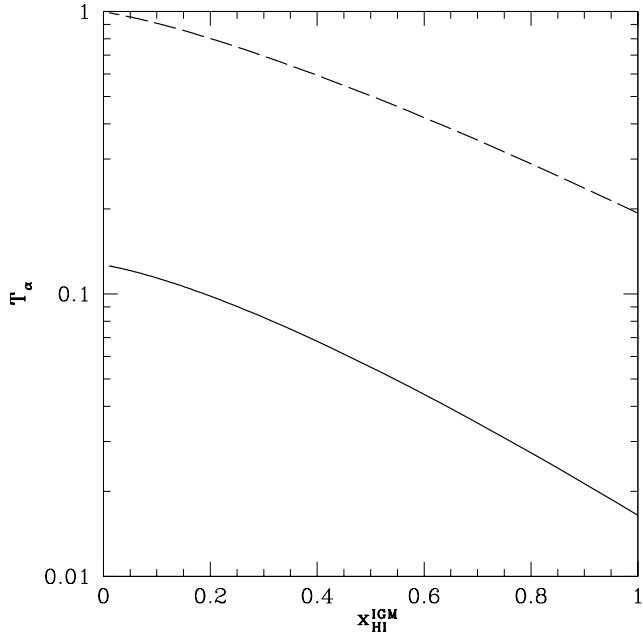}
  \caption{Ratio of observed to intrinsic \lya\ line flux as a
  function of $\xigm$.  The solid curve assumes the intrinsic line
  profile described in Section~\ref{sec:inline}.  The long-dashed
  curve assumes the intrinsic \lya\ line is concentrated $360~\kms$
  redward of the systemic velocity (see Section~\ref{sec:galwind}).
  Both curves are for the fiducial galaxy.}
  \label{fig:xigmhital_fx01}
\end{figure}

Once reliable estimates of $T_\alpha$ are available, they can be used
to estimate the neutral fraction of the IGM in the interesting regime
of $\xhi\sim0.5$.  Figure~\ref{fig:xigmhital_fx01} illustrates
$T_\alpha$ as a function of the model parameter $\xigm$ (see
Section~\ref{sec:igmiondep}), for the fiducial galaxy.  The solid
curve was calculated assuming a Doppler profile intrinsic \lya\ line
centered at the galaxy's systemic redshift, and the long-dashed curve
was calculated assuming the intrinsic \lya\ line is concentrated
$360~\kms$ redward of the systemic velocity.  While an accurate
measurement of $T_\alpha$ can constrain the IGM neutral fraction, this
is only possible with reasonable knowledge of the redshift of the
intrinsic \lya\ line compared to the systemic velocity of the galaxy.

\subsection{Model assumptions}
\label{sec:modass}

Now we return to a discussion of the assumptions made in this paper
and their possible consequences on our conclusions.  First we discuss
the possible effects of dust in a \lya\ emitting galaxy.  If dust
absorbs ionizing photons, then the star-formation rate in our models
can simply be rescaled to take that into account.  Alternately, dust
may extinct the \lya\ line and UV continuum almost equally, but
extinct the Balmer lines considerably less.  In this case the \lya\
line extinction can be estimated and corrected for.  Dust may also
affect the UV continuum differently from the recombination emission,
in the event that dust is distributed differently with respect to
stars and \hii\ regions.  In this case the extinction value measured
by the Balmer lines alone (with an extinction curve) may be used to
correct the \lya\ line, assuming that resonant photons are extincted
no differently than non-resonant photons.

Dust extinction can cause great ambiguity in the interpretation of
\lya\ lines, even when UV continuum and Balmer lines are measured,
because resonant \lya\ photons may be extincted differently from
non-resonant UV continuum and Balmer line photons.  Dust may extinct
\lya\ photons preferentially over UV continuum photons, due to the
extra path-length through the galaxy that resonantly scattered photons
travel before escaping.  Certain gas and dust geometries, however, may
mitigate or even reverse that effect \citep{neu91}.  There is no
direct observable to measure this possible effect, except that an
extinction-corrected SFR estimated from Balmer lines does place an
upper limit on the intrinsic \lya\ line flux.

More fundamental assumptions are associated with our model of the IGM
around \lya\ emitting galaxies.  We assumed the presence of an
accretion shock at the virial radius of the halo containing the galaxy
\citep[see also][]{bar03b}.  If these galaxies have no accretion
shock, as suggested by \citet{bir03}, then our conclusions are greatly
changed.  It is the presence of the accretion shock that generates the
characteristic wavelength scale where line transmission begins (see
Section~\ref{sec:igmmoddep}).  In the absence of a virial shock, the
infalling gas will eventually shock when it reaches the galaxy.  The
observed \lya\ line would be suppressed blueward of that velocity,
implying only intrinsic \lya\ lines that were very redshifted with
respect to the systemic redshift of the galaxy could be observed.
Another consequence of the lack of a virial shock is that \lya\
emission from the cooling of the IGM gas would be located close to the
galaxy \citep{bir03}, thus any observed \lya\ line may have a
contribution from contribution from cooling radiation as well as
star-formation powered \lya.  If future observations show an extremely
redshifted \lya\ lines at high redshifts, this may be indicative of
cooling radiation from IGM gas falling unshocked down to the center of
a halo.

In Section~\ref{sec:igmmodel} we discussed our assumptions about the
density distribution of the IGM around the \lya\ emitting galaxy.  We
calculate the ensemble-averaged IGM density distribution; even for the
spherically-averaged density profile, deviations from this average are
expected from galaxy to galaxy.  Additionally, we estimated that
structure near the emitting galaxy has only a small effect on the
density distribution of the IGM.  However, we did not address the
effect of collapsed and collapsing structures on the velocity field of
the IGM, which is crucial (more so than IGM density) to our predicted
line profiles.  Numerical simulations may be used to estimate both the
variance in the IGM structure from galaxy to galaxy, and also the
density and velocity distributions of the IGM along the line of sight
toward a galaxy accounting for substructure.

\section{Summary}
\label{sec:summ}

We have calculated observed \lya\ line profiles for high-redshift
galaxies, with a focus on the prospects of using observed \lya\ lines
to probe the ionization of the universe at $z>6.5$.  Toward this end
we have investigated many of the parameters that influence the
observed \lya\ line from a high-redshift galaxy.  We conclude that
future measurements with \textit{JWST} may permit the estimation of
IGM neutral fractions at $z>6.5$, but current $z=6.5$ data do not yet
place a strong constraint on the neutral fraction at that redshift if
we allow consideration of galactic winds.  If we take at face value
our estimates of the integrated \lya\ transmission of the two $z=6.5$
galaxies for which estimation is possible, and assume our simplest
model (with no winds), we would conclude that the IGM neutral fraction
is $\xhi\la0.1$ at $z=6.5$.

Interpretation of an observed \lya\ line alone (or even with a
measurement of the UV continuum) is extremely difficult, due to the
host of parameters that influence it.  However there is hope that with
the rest-frame optical spectra that \textit{JWST} will provide for
high-redshift galaxies, \lya\ emission lines will yet prove fruitful
for constraining the reionization history of the universe at $z>6.5$

\section*{Acknowledgments}
We acknowledge Kurt Adelberger for suggesting the IGM model used here.
We thank Marc Kamionkowski, Avi Loeb, and Alice Shapley for helpful
conversations.

\bsp

\label{lastpage}

\end{document}